\documentclass[showpacs,prl,aps,superscriptaddress,preprintnumbers,nofootinbib]{revtex4-2}
\usepackage[T1]{fontenc}
\usepackage{textcomp}
\usepackage[latin9]{inputenc}
\usepackage{colortbl}
\usepackage{mathtools}
\usepackage{amsmath}
\usepackage{amssymb}
\usepackage{graphicx}
\usepackage[pdfusetitle,
 bookmarks=true,bookmarksnumbered=false,bookmarksopen=false,
 breaklinks=false,pdfborder={0 0 1},backref=false,colorlinks=false]
 {hyperref}

\makeatletter

\providecommand{\tabularnewline}{\\}

\providecommand{\tabularnewline}{\\}

\@ifundefined{textcolor}{}{%
 \definecolor{BLACK}{gray}{0}
 \definecolor{WHITE}{gray}{1}
 \definecolor{RED}{rgb}{1,0,0}
 \definecolor{GREEN}{rgb}{0,1,0}
 \definecolor{BLUE}{rgb}{0,0,1}
 \definecolor{CYAN}{cmyk}{1,0,0,0}
 \definecolor{MAGENTA}{cmyk}{0,1,0,0}
 \definecolor{YELLOW}{cmyk}{0,0,1,0}
}

\usepackage{epsfig}
\usepackage{bm}
\usepackage{array}
\usepackage{amsfonts}
\usepackage{epstopdf}
\usepackage{mathrsfs}
\usepackage{color}
\usepackage{pifont}
\usepackage{relsize}
\def\slashchar#1{\setbox0=\hbox{$#1$}     		
   \dimen0=\wd0                                 	
   \setbox1=\hbox{/} \dimen1=\wd1               	
   \ifdim\dimen0>\dimen1                        	
      \rlap{\hbox to \dimen0{\hfil/\hfil}}      	
      #1                                        	
   \else                                        	
      \rlap{\hbox to \dimen1{\hfil$#1$\hfil}}   	
      /                                         	
   \fi}

\newcommand{\del}{\boldsymbol{\Delta}}

\makeatother

\begin{document}
\title{Constraints on the odderon amplitude in the CGC framework}
\author{M. Roa}
\email{michael.roa@upla.cl}

\affiliation{Departamento de F\'isica, Universidad T\'ecnica Federico Santa Mar\'ia,
Casilla 110-V, Valpara\'iso, Chile}
\affiliation{Facultad de Ingenier\'ia, Laboratorio DataScience, Universidad de Playa
Ancha, ~~\\
 Leopoldo Carvallo 270, Valpara\'iso, Chile~~~~~~ ~~\\
}
\author{M. Siddikov}
\email{marat.siddikov@usm.cl}

\affiliation{Departamento de F\'isica, Universidad T\'ecnica Federico Santa Mar\'ia,
Casilla 110-V, Valpara\'iso, Chile}
\affiliation{Centro Cient\'ifico - Tecnol\'ogico de Valpara\'iso, Casilla 110-V, Valpara\'iso,
Chile}
\author{Y. Gentile}
\email{yanil.garrido@alumnos.upla.cl}

\affiliation{Facultad de Ingenier\'ia, Laboratorio DataScience, Universidad de Playa
Ancha, ~~\\
 Leopoldo Carvallo 270, Valpara\'iso, Chile~~~~~~ ~~\\
}
\author{I. Zemlyakov}
\email{izemlyakov@usm.cl}

\affiliation{Departamento de F\'isica, Universidad T\'ecnica Federico Santa Mar\'ia,
Casilla 110-V, Valpara\'iso, Chile}
\affiliation{Instituto de F\'isica, Pontificia Universidad Cat\'olica de Valpara\'iso,
Av. Brasil 2950, Valpara\'iso, Chile}
\date{\today}

\begin{abstract}
In this manuscript we analyzed the elastic $pp$ and $p\bar{p}$ cross-section
in the Color Glass Condensate framework, treating them as a dilute-dense
system, and derived the phenomenological constraints (upper bounds)
on the odderon-mediated part of the dipole scattering amplitude $\mathcal{O}(Y,\,\boldsymbol{r},\,\boldsymbol{b})$.
For our analysis we used the experimental data available from TOTEM-D0
and ISR collaborations, and constructed a combination of the cross-sections
which is directly proportional to the odderon-mediated part of the
amplitude. This allows to suppress possible uncertainties associated
with charge-parity even part of the amplitude. We analyzed two phenomenological
parametrizations of odderons and demonstrated that after minor adjustments
to the global normalization, they can describe the experimental data
reasonably well, although due to large experimental uncertainties
the odderon amplitude remains loosely constrained.
\end{abstract}
\maketitle

\section{Introduction}

The odderon, a charge parity odd ($C$-odd) counterpart of the well-established
Pomeron exchange, remains one of the most elusive objects in the theory
of strong interactions. Originally predicted in the 1970s within the
framework of Regge theory~\cite{Lukaszuk:1973nt,Gauron:1992zc},
it is interpreted in the language of Quantum Chromodynamics (QCD)
as the exchange of an odd number of reggeized gluons in a color singlet
configuration in the $t$-channel~\cite{Kovchegov:2012mbw,Ewerz:2003xi,Bartels:1999yt}.
In the dilute (perturbative) systems, the odderon arises as a coherent
exchange of three interacting reggeized gluons in the $t$-channel,
and its amplitude should satisfy the Bartels-Kwiecinski-Praszalowicz
(BKP) equation~\cite{Bartels:1980pe,Kwiecinski:1980wb}, written
for the special case of three reggeized gluons. Beyond that limit,
the odderon has been explored in non-perturbative approaches such
as the Stochastic Vacuum Model~\cite{Dosch:1988ha} and holographic
QCD~\cite{Brower:2008cy}, where it emerges from the exchange of
specific string modes in the dual gravitational background. In the
Color Glass Condensate (CGC) framework, the odderon admits a natural
description in terms of correlators of Wilson lines with definite
symmetry under charge conjugation, and its evolution can be studied
via extensions of the BK-JIMWLK equations~\cite{Kovner:2005qj,Kovchegov:2003dm,Hatta:2005as}.

Since the odderon exchange is an essentially nonperturbative phenomenon,
at present the amplitude of the odderon-mediated scattering cannot
be predicted unambiguously from the first principles and should be
extracted from the experimental data. Phenomenologically, the odderon
can manifest itself in observables that are sensitive to the difference
between particle and antiparticle scattering~\cite{TOTEM:2017asr,TOTEM:2017sdy,TOTEM:2018hki},
or in exclusive processes which require exchange of $C$-parity in
$t$-channel~\cite{Benic:2023ybl}. However, the experimental identification
of the odderons remains challenging due to significant contributions
from various pomeron-mediated mechanisms that can mimic a weak odderon
signal or provide sizable backgrounds. At present the most compelling
evidence in favor of odderons comes from a recent comparison of the
differential cross-sections for proton-proton and proton-antiproton
elastic scattering at the LHC and Tevatron~\cite{Martynov:2017zjz,D0:2020tig,D0:2012erd,TOTEM:2018psk}.
However, the estimates of the statistical significance of this signal
vary significantly due to experimental and theoretical ambiguities
(see a short overview of recent developments in~\cite{Csorgo:2024dvr,ATLAS:2022mgx}).
For odderons these uncertainties are especially pronounced because
in the cross-section it contributes commingled with a large $C$-even
contribution, so the small errors of the latter can drastically change
the estimates of the odderon amplitude and even cast doubt on its
existence. This motivates the construction of the observables designed
to be exclusively sensitive to odderon exchange, free from contamination
by $C$-even backgrounds. On the other hand, the estimates of the
odderon signal at the future experiments are based on phenomenological
parametrizations~\cite{Yao:2018vcg,Benic:2023ybl} that include an
arbitrary parameter which controls the magnitude of odderon. In this
manuscript we try to find the upper limit on the odderon amplitude
in the CGC framework and also propose a new observable (combination
of cross-sections) that could be more sensitive to the odderon contributions.

The paper is structured as follows. In the following section~\ref{sec:Theor}
we briefly summarize the main theoretical tools used for description
of the proton structure in high energy collisions. In Section~\ref{sec:amplitude}
we analyze in detail the elastic proton-proton collision scattering,
discuss the role of odderons and introduce the observable that allows
to extract the contributions of pure odderons. Later in the section~\ref{subsec:data}
we perform phenomenological analysis and compare theoretical predictions
for this observable with existing experimental data. We also analyze
upper limits for the odderon amplitude which are compatible with data.
Finally in Section~\ref{sec:Conclusions} we draw conclusions.

\section{Theoretical framework}

\label{sec:Theor}

\subsection{Structure of the proton and its description in high energy processes}

\label{subsec:harmonic}

\begin{figure}
\includegraphics[width=8.4cm]{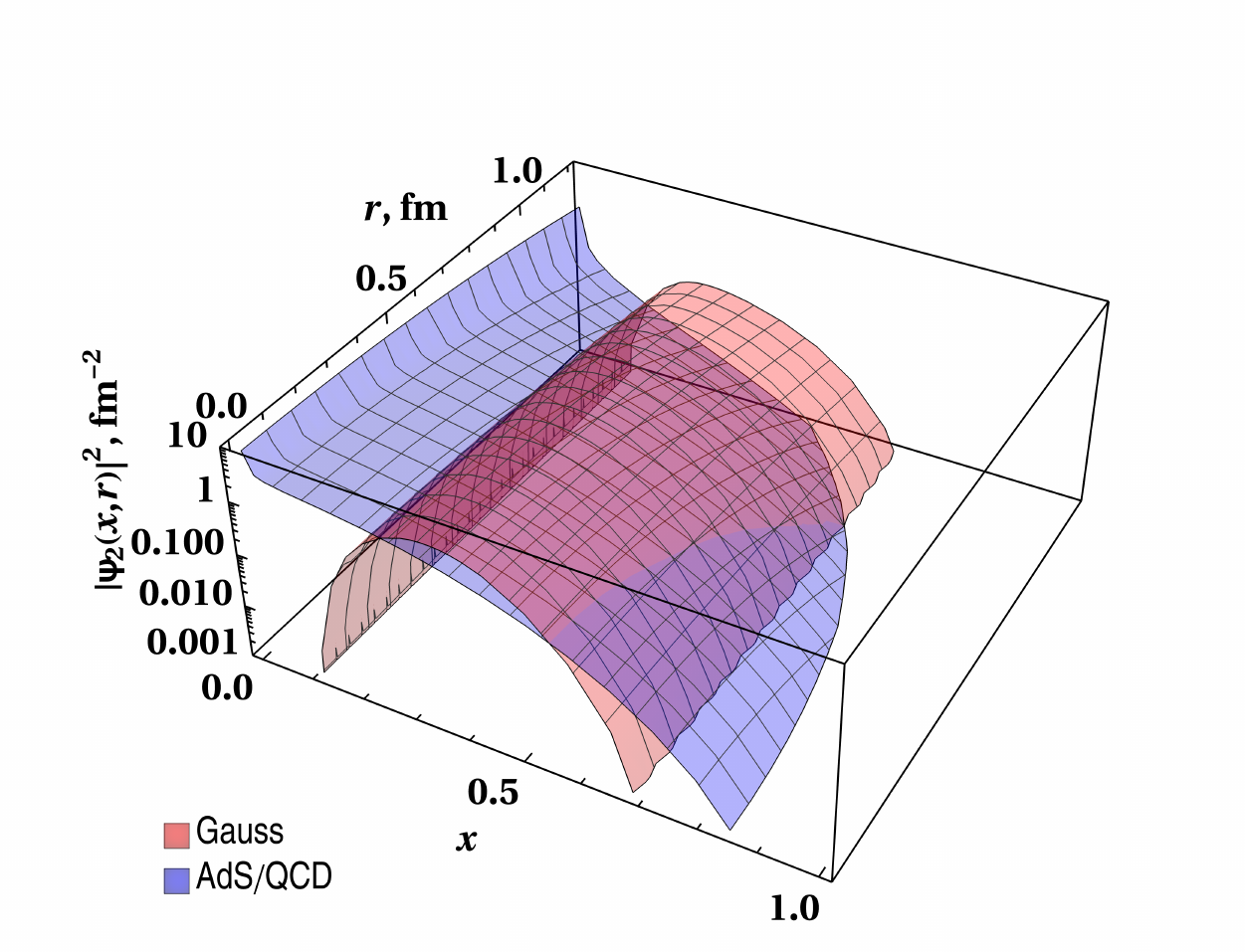}\includegraphics[width=8.4cm]{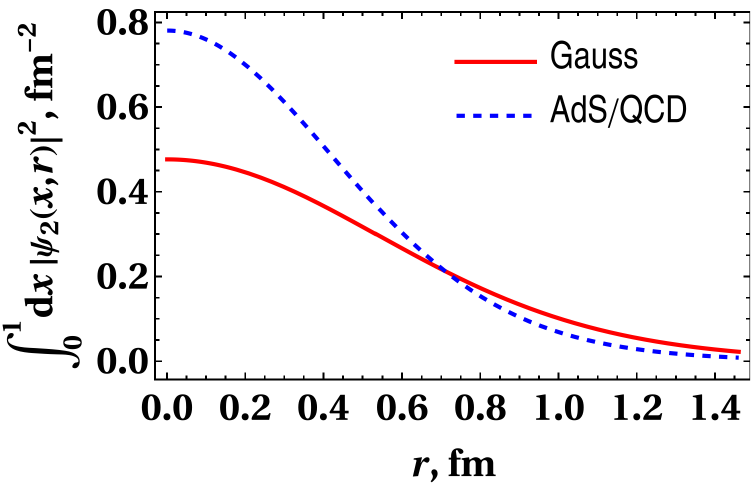}

\caption{(Color online) Left: Comparison of the wave functions~(\ref{eq:dIgauss})
and~(\ref{LFWFb}). Right: comparison of the $x$-integrated wave
functions~(\ref{eq:dIgauss}) and~(\ref{LFWFb}). See the text for
more details.}
\label{fig:WF}
\end{figure}

In general the elementary particle is a sophisticated ensemble of
Fock states with different number of quarks and gluons. However, we
expect that the dominant contributions to many processes comes from
the valence three-quark component of the proton. The light-cone wave
function of the latter is given by~\cite{Lepage:1980fj,Brodsky:1997de,Dumitru:2019qec}:

\begin{eqnarray}
|P\rangle & \approx & \frac{1}{\sqrt{6}}\int\frac{dx_{1}dx_{2}dx_{3}}{\sqrt{x_{1}x_{2}x_{3}}}\delta\left(1-x_{1}-x_{2}-x_{3}\right)\int\frac{d^{2}k_{1}d^{2}k_{2}d^{2}k_{3}}{(16\pi^{3})^{3}}\,16\pi^{3}\delta\left(\boldsymbol{k}_{1}+\boldsymbol{k}_{2}+\boldsymbol{k}_{3}\right)\nonumber \\
 & \times & \psi_{3}\left(x_{1},\boldsymbol{k}_{1};\,x_{2},\boldsymbol{k}_{2};\,x_{3},\boldsymbol{k}_{3}\right)\sum_{i_{1},i_{2},i_{3}}\epsilon_{i_{1}i_{2}i_{3}}|p_{1},i_{1},f_{1};\,p_{2},i_{2},f_{2};\,p_{3},i_{3},f_{3}\rangle~.\label{DUMWF}
\end{eqnarray}
where $P^{\mu}=\left(P^{+},m^{2}_{N}/2P^{+},\,\boldsymbol{0}_{\perp}\right)$
is the 4-momentum of the proton, and $\psi_{3}(x_{i},\boldsymbol{k}_{i})$
is the universal (process-independent) nonperturbative light-front
wave function of the three-quark Fock state. In Eq~(\ref{DUMWF})
the variables $x_{i}$ represent the longitudinal momentum carried
by each quark, and $\boldsymbol{k}_{i}$ characterize the transverse
momenta of the internal motion of the quarks in the transverse plane.
The Levi-Civita tensor $\epsilon_{i_{1}i_{2}i_{3}}$ reflects overall
antisymmetry with respect to the color indices of individual quarks
in the color singlet proton. Finally, the indices $f_{1},f_{2},f_{3}$
correspond to flavor labels of individual quarks. In our approach,
we use this valence Fock state as the starting point for modeling
the exclusive observables. The wave function is normalized such that
the proton state satisfies the usual light-front normalization condition
\begin{eqnarray}
\langle K|P\rangle & = & 16\pi^{3}\,P^{+}\delta(P^{+}-K^{+})\,\delta(\boldsymbol{P}_{\perp}-\boldsymbol{K}_{\perp})~,\label{eq:ProtonNorm1}
\end{eqnarray}
which for the 3-quark system translates into 
\begin{eqnarray}
\int{dx_{1}dx_{2}dx_{3}}\,\delta\left(1-x_{1}-x_{2}-x_{3}\right)\int\frac{{d^{2}k_{1}d^{2}k_{2}d^{2}k_{3}}}{(16\pi^{3})^{3}}\,(16\pi^{3})\,\delta\left(\boldsymbol{k}_{1}+\boldsymbol{k}_{2}+\boldsymbol{k}_{3}\right)\,\left|\psi_{3}\right|^{2}=1~.\label{eq:Norm_psi3}
\end{eqnarray}
For phenomenological analysis frequently the wave function is chosen
in the Gaussian form~\cite{Brodsky:1994fz},

\begin{eqnarray}
\psi^{(G)}_{3}\left(x_{1},\boldsymbol{k}_{1};x_{2},\boldsymbol{k}_{2};x_{3},\boldsymbol{k}_{3}\right) & = & N\exp(-{\cal M}^{2}/2\beta^{2}),\qquad{\cal M}^{2}=\sum^{3}_{i=1}\frac{\boldsymbol{k}^{2}_{\perp i}+m^{2}}{x_{i}}\label{eq:Gauss}
\end{eqnarray}
where $N$ is the normalization constant. The mass of the constituent
quark $m$ and the parameter $\beta$ are treated as free parameters
fitted to the electroweak properties of the baryon octet as explained
in Ref.~\cite{Schlumpf:1993rm,Schlumpf:1992vq}. The main advantage
of the parametrization~(\ref{eq:Gauss}) is its simplicity, which
allows to get analytic results for the configuration space wave function.
At very high energy processes, only one of the three quarks participates
(interacts with other partons), so the 3-quark system may be reduced
to a simpler quark-diquark system, which is characterized by the light-cone
fraction $x$ carried by the active quark and the effective separation
$\boldsymbol{r}$ between the quark and the center of mass of two
spectators. This approximation for the three-quark system was proposed in the early papers on the baryon structure in QCD~\cite{GellMann:1964ewy,Ida:1966ev,Lichtenberg:1967zz}
and since that time has been constantly used as a basis for various
phenomenological models of the proton structure. The corresponding
(squared of) wave function of the quark-diquark system is given by
\begin{equation}
\left|\psi_{2}\left(x,\,\boldsymbol{r}\right)\right|^{2}=\int dx_{2}\int\frac{{d^{2}k_{1}d^{2}k_{2}}}{(16\pi^{3})^{2}}e^{i\boldsymbol{k}_{1}\cdot\boldsymbol{r}}\,\left|\psi_{3}\left(x,\boldsymbol{k}_{1};\,x_{2},\boldsymbol{k}_{2};\,x_{3}=1-x-x_{2},\,\boldsymbol{k}_{3}=-\boldsymbol{k}_{1}-\boldsymbol{k}_{2}\right)\right|^{2}.
\end{equation}
For the Gaussian parametrization~(\ref{eq:Gauss}), we may obtain
\begin{equation}
\left|\psi^{(G)}_{2}\left(x,\,\boldsymbol{r}\right)\right|^{2}=\frac{N^{2}\beta^{4}\,x}{256\pi^{4}}\int^{1-x_{1}}_{0}dx_{2}\,x_{2}(1-x-x_{2})\exp\left(-\frac{\beta^{2}\boldsymbol{r}^{2}x(1-x)}{4}-\frac{m^{2}\left((x+x_{2})(1-x_{2})-x^{2}\right)}{\beta^{2}\,xx_{2}(1-x-x_{2})}\right)\label{eq:dIgauss}
\end{equation}
For further numerical estimates, we used the values of parameters
$m=0.26$ GeV and $\beta=1.1$ GeV provided in~\cite{Schlumpf:1993rm,Schlumpf:1992vq},
though we had to double the value of the parameter $\beta$ in order
to match the experimental root-mean-square (rms) radius of the proton~\cite{PDG}
\begin{equation}
\left\langle r^{2}_{p}\right\rangle =\frac{\int^{1}_{0}dx\int d^{2}r\,\,r^{2}\,\left|\psi^{(G)}_{2}(x,\,\boldsymbol{r})\right|^{2}}{\int^{1}_{0}dx\int d^{2}r\,\,\left|\psi^{(G)}_{2}(x,\,\boldsymbol{r})\right|^{2}}\approx\left(0.83\,\text{fm}\right)^{2}.\label{eq:r2ms}
\end{equation}
While the function~(\ref{eq:dIgauss}) provides a reasonable estimate
for the variables dominated by the valence quarks, it is not universal.
In order to assess the model dependence introduced by the choice of
the wave function, we will also use in parallel the AdS/QCD-based
parametrization suggested in Ref.~\cite{deTeramond:2018ecg},

\begin{equation}
\psi^{({\rm AdS/QCD})}_{2}\left(x,\,\boldsymbol{r}\right)=\frac{1}{2\sqrt{\pi}}\sqrt{\frac{q_{\tau}(x)}{f(x)}}(1-x)\exp\left[-\frac{(1-x)^{2}}{8f(x)}\,\boldsymbol{r}^{2}\right],\label{LFWFb}
\end{equation}
where the parameter $\tau$ denotes the so-called twist of the Fock
component. The functions $q_{\tau}(x)$ and $f(x)$ that control the
longitudinal and transverse shape, are given by~\cite{Brodsky:2014yha,Brodsky:2016yod}
\begin{align}
q_{\tau}(x) & =\frac{1}{N_{\tau}}\left(1-w(x)\right)^{\tau-2}w(x)^{-1/2}w'(x),\qquad f(x)=\frac{1}{4\Lambda}\log\left(\frac{1}{w(x)}\right),\label{qx}\\
w(x) & =x^{1-x}e^{-a(1-x)^{2}},\,\,\,a=0.531\pm0.037,\,\,\sqrt{\Lambda}=0.548~\text{GeV}.
\end{align}
Similar to~(\ref{eq:dIgauss}), the parametrization~(\ref{LFWFb})
exhibits a Gaussian fall-off at large distances, reflecting the confinement,
although with different $x$-dependent slope. We checked that this
parametrization gives acceptable value of the root-mean-square (rms)
radius $\sqrt{\left\langle r^{2}\right\rangle }\approx0.7~\text{fm}$,
which is approximately twenty percent smaller than~(\ref{eq:r2ms}).
In the Figure~\ref{fig:WF} we have shown the wave functions~(\ref{eq:dIgauss})
and~(\ref{LFWFb}), together with their first moments $\int dx\left|\psi_{2}(x,r)\right|^{2}$
which will be needed later. In what follows we will use for our estimates
both wave functions, considering the difference between them as a
tentative estimate of the size of nonperturbative effects.

\subsection{Dipole scattering in the CGC framework}

\label{sec:odderon}

\begin{figure}
\begin{tabular}{cc}
\includegraphics[width=8cm]{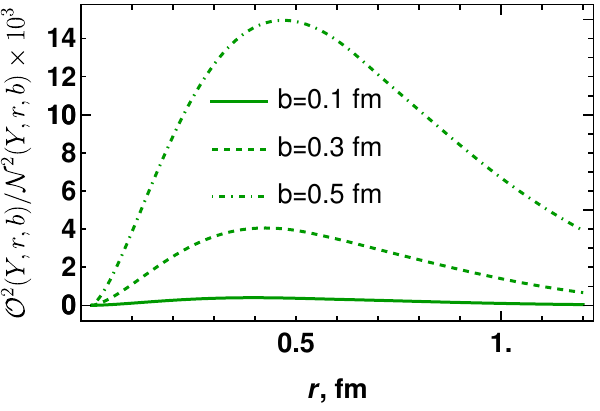} & \includegraphics[width=8cm]{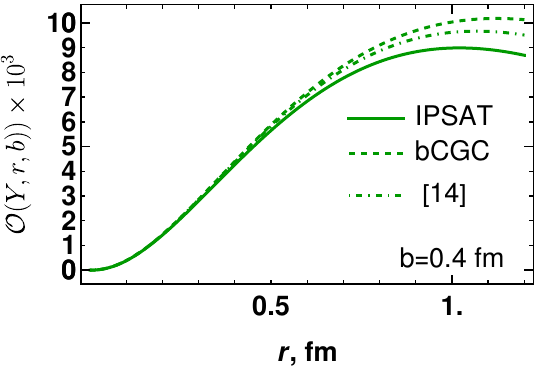}\tabularnewline
\end{tabular}

\caption{\textbf{\emph{}}Left: The ratio of the $C$-odd and $C$-even terms
$\sim\mathcal{O}^{2},\mathcal{N}^{2}$ that appear in the right-hand
side of~(\ref{eq:Pom}). For the sake of definiteness, we used the
parametrization of odderons and pomerons from~\cite{Benic:2023ybl},
but we checked that the parametrization from~\cite{Hatta:2005as,Yao:2018vcg}
also gives small numbers which do not exceed a few per cent. This
plot clearly demonstrates that the odderons are negligible for the
evolution of the $C$-even dipole amplitude. Right: The dependence
of the odderon evolution on the parametrization of the dipole amplitude.
At the initial scale we use parametrization from~\cite{Benic:2023ybl}
for the odderon amplitude, and parametrizations from~\cite{Benic:2023ybl,Rezaeian:2013tka,Rezaeian:2012ji}
for the $C$-even amplitude $\mathcal{N}\left(Y,\,\boldsymbol{r},\,\boldsymbol{b}\right)$
(the latter three are fitted to the experimental data from HERA~\cite{Aaron:2009aa}).
The plot represents the amplitude $\mathcal{O}\left(Y,\,\boldsymbol{r},\,\boldsymbol{b}\right)$
at the final scale (rapidity difference $\Delta Y=Y_{{\rm fin}}-Y_{{\rm ini}}=\ln\left(10^{3}\right)$).
In the region $r\lesssim1\,{\rm fm}$, which gives the dominant contribution
to the cross-section, the uncertainty due to choice of the dipole
amplitude does not exceed a few per cent.}\label{fig:Assessment}
\label{fig:odderon_Benic-1-1}
\end{figure}

The CGC framework describes the high-energy scattering of the dilute-dense
system in the eikonal approximation. In this picture the interaction
of each parton of the (dilute) projectile with the target is encoded
in the universal Wilson lines $U(\boldsymbol{x}_{\perp})$~ defined
as 
\begin{equation}
U\left(\boldsymbol{x}_{\perp}\right)=P\exp\left(ig\int dx^{-}A^{+}_{a}\left(x^{-},\,\boldsymbol{x}_{\perp}\right)t^{a}\right),\label{eq:Wilson}
\end{equation}
where we use notation $\boldsymbol{x}_{\perp}$ for the impact parameter
of the parton, $A^{\mu}_{a}$ for the gluonic field of the target,
and $t_{a}$ are the usual color group generators in the irreducible
representation which corresponds to the parton ( $\boldsymbol{3}$,
$\boldsymbol{\bar{3}}$ or $\boldsymbol{8}$ for quark, antiquark
or gluon, respectively). The interaction of the quark-antiquark (or
quark-diquark) system in this approach are given by the two-point
correlator~\cite{Kovchegov:2003dm,Hatta:2005as} 
\begin{align}
\frac{1}{N_{c}}\left\langle {\rm tr}\left(U\left(\boldsymbol{x}_{q}\right)U^{\dagger}\left(\boldsymbol{x}_{\bar{q}}\right)\right)\right\rangle _{Y} & =1-\mathcal{N}\left(Y,\,\boldsymbol{r},\,\boldsymbol{b}\right)+i\mathcal{O}\left(Y,\,\boldsymbol{r},\,\boldsymbol{b}\right),\label{eq:dipo}\\
\boldsymbol{r} & =\boldsymbol{x}_{q}-\boldsymbol{x}_{\bar{q}},\quad\boldsymbol{b}=z\boldsymbol{x}_{q}+(1-z)\boldsymbol{x}_{\bar{q}}\nonumber 
\end{align}
where the angular bracket $\left\langle ...\right\rangle _{Y}$ in
the left-hand side of~(\ref{eq:dipo}) stands for averaging over
all the gluonic field configurations of the target, $Y$ is the rapidity
of the dipole, $\boldsymbol{r}$ is the separation between the quarks,
and $\boldsymbol{b}$ is the impact parameter. The functions $\mathcal{N},\,\mathcal{O}$
correspond to $C$-even and $C$-odd parts of the dipole scattering
amplitude, respectively, and are frequently referred to as ``pomeron''
and ``odderon'' contributions. These functions have different parity
under permutation of the quark and antiquark coordinates, namely the
function $\mathcal{O}$ changes its sign under $\boldsymbol{r}\to-\boldsymbol{r}$,
whereas the function $\mathcal{N}$ does not change under this transformation.
This implies that the function $\mathcal{O}$ should have nontrivial
dependence on the orientation of the vector $\boldsymbol{r}$ with
respect to some external vector (e.g. impact parameter $\boldsymbol{b}$
or spin $\boldsymbol{S}$). The rapidity dependence of the functions
$\mathcal{N},\,\mathcal{O}$ should satisfy the Balitsky-Kovchegov
evolution equations~\cite{Kovner:2005qj,Kovchegov:2003dm,Hatta:2005as}
\begin{align}
\frac{\partial\mathcal{N}\left(Y,\,\boldsymbol{r},\,\boldsymbol{b}\right)}{\partial Y} & =\int_{\boldsymbol{r}_{1}}\mathcal{K}\left(\boldsymbol{r},\,\boldsymbol{r}_{1}\right)\left[\mathcal{N}\left(Y,\,\boldsymbol{r}_{1},\,\boldsymbol{b}\right)+\mathcal{N}\left(Y,\,\boldsymbol{r}_{2},\,\boldsymbol{b}\right)-\mathcal{N}\left(Y,\,\boldsymbol{r},\,\boldsymbol{b}\right)\right.\label{eq:Pom}\\
 & \qquad-\left.\mathcal{N}\left(Y,\,\boldsymbol{r}_{1},\,\boldsymbol{b}\right)\mathcal{N}\left(Y,\,\boldsymbol{r}_{2},\,\boldsymbol{b}\right)+\mathcal{O}\left(Y,\,\boldsymbol{r}_{1},\,\boldsymbol{b}\right)\mathcal{O}\left(Y,\,\boldsymbol{r}_{2},\,\boldsymbol{b}\right)\right],\nonumber \\
\frac{\partial\mathcal{O}\left(Y,\,\boldsymbol{r},\,\boldsymbol{b}\right)}{\partial Y} & =\int_{\boldsymbol{r}_{1}}\mathcal{K}\left(\boldsymbol{r},\,\boldsymbol{r}_{1}\right)\left[\mathcal{O}\left(Y,\,\boldsymbol{r}_{1},\,\boldsymbol{b}\right)+\mathcal{O}\left(Y,\,\boldsymbol{r}_{2},\,\boldsymbol{b}\right)-\mathcal{O}\left(Y,\,\boldsymbol{r},\,\boldsymbol{b}\right)\right.\label{eq:Odd}\\
 & \qquad-\left.\mathcal{N}\left(Y,\,\boldsymbol{r}_{1},\,\boldsymbol{b}\right)\mathcal{O}\left(Y,\,\boldsymbol{r}_{2},\,\boldsymbol{b}\right)-\mathcal{O}\left(Y,\,\boldsymbol{r}_{1},\,\boldsymbol{b}\right)\mathcal{N}\left(Y,\,\boldsymbol{r}_{2},\,\boldsymbol{b}\right)\right],\nonumber 
\end{align}
where $\boldsymbol{r}_{2}=\boldsymbol{r}-\boldsymbol{r}_{1}$, and
$\mathcal{K}\left(\boldsymbol{r},\,\boldsymbol{r}_{1}\right)$ in~(\ref{eq:Pom},~\ref{eq:Odd})
is the standard BFKL kernel with running coupling corrections, 
\begin{equation}
\mathcal{K}\left(\boldsymbol{r},\,\boldsymbol{r}_{1}\right)=\frac{\alpha_{s}\left(\boldsymbol{r}\right)N_{c}}{2\pi^{2}}\left[\frac{\boldsymbol{r}^{2}}{\boldsymbol{r}^{2}_{1}\boldsymbol{r}^{2}_{2}}+\frac{1}{\boldsymbol{r}^{2}_{1}}\left(\frac{\alpha_{s}\left(\boldsymbol{r}_{1}\right)}{\alpha_{s}\left(\boldsymbol{r}_{2}\right)}-1\right)+\frac{1}{\boldsymbol{r}^{2}_{2}}\left(\frac{\alpha_{s}\left(\boldsymbol{r}_{2}\right)}{\alpha_{s}\left(\boldsymbol{r}_{1}\right)}-1\right)\right].
\end{equation}
The dependence on the impact parameter $\boldsymbol{b}$ in~(\ref{eq:Pom},~\ref{eq:Odd})
appears only parametrically, which implies that the evolution can
be done independently for each $\boldsymbol{b}$. Due to the above-mentioned
different symmetry properties of the pomerons and odderons, the dipole
amplitude $\mathcal{N}\left(Y,\,\boldsymbol{r},\,\boldsymbol{b}\right)$
saturates (can't exceed unity) at very high energy, whereas odderons
tend to decrease with energy~\cite{Contreras:2020lrh}. While the function $\mathcal{N}\left(Y,\,\boldsymbol{r},\,\boldsymbol{b}\right)$
is reasonably well understood from the phenomenological fits of the
Deep Inelastic Scattering (at least for small size dipoles), the function
$\mathcal{O}\left(Y,\,\boldsymbol{r},\,\boldsymbol{b}\right)$ at
present is poorly constrained, although remains significantly smaller
than $\mathcal{N}$ in all the available parametrizations. From the
left panel of the Figure~\ref{fig:Assessment} we can see that
the ratio of the $C$-odd to $C$-even components of the dipole scattering
amplitude $\left|\mathcal{O}\right|^{2}/\left|\mathcal{N}\right|^{2}$
does not exceed a few per cent. For this reason, the last term in
the second line of~(\ref{eq:Pom}) may be safely disregarded, so
the equation (\ref{eq:Pom}) decouples from odderons. This justifies
the odderon-independent analysis of the $C$-even dipole amplitude,
and allows to fix the latter with very good precision using available
precise data from HERA~\cite{Aaron:2009aa}.
The remaining equation (\ref{eq:Odd}) at fixed $\mathcal{N}\left(Y,\,\boldsymbol{r},\,\boldsymbol{b}\right)$
turns into a linear homogeneous integro-differential equation with
respect to $\mathcal{O}$. In order to understand how much the uncertainty
of the dipole amplitude $\mathcal{N}\left(Y,\,\boldsymbol{r}_{1},\,\boldsymbol{b}\right)$
affects the odderon evolution, we evaluated numerically the evolution
of the odderon amplitude from~\cite{Benic:2023ybl} over a large
interval $\Delta Y=\ln\left(10^{3}\right)$ using three different
parametrization of $\mathcal{N}$ from~~\cite{Benic:2023ybl,Rezaeian:2013tka,Rezaeian:2012ji}
(see Section~\ref{subsec:data} for more details). As we can see
from the right panel of the Figure~\ref{fig:Assessment}, the uncertainty
of the dipole amplitude $\mathcal{N}\left(Y,\,\boldsymbol{r},\,\boldsymbol{b}\right)$
has only minor influence on the result of evolution, and does not
exceed a few per cent in the region $r\lesssim1\,{\rm fm}$ which
gives the dominant contribution. Since the Eq.~(\ref{eq:Odd}) is
a linear homogeneous integro-differential equation with respect to
$\mathcal{O}$, formally its general solution is defined up to rescaling
transformation 
\begin{equation}
\mathcal{O}\left(Y,\,\boldsymbol{r},\,\boldsymbol{b}\right)\to\lambda\,\mathcal{O}\left(Y,\,\boldsymbol{r},\,\boldsymbol{b}\right),\quad\lambda={\rm const}.\label{eq:lambda}
\end{equation}
where $\lambda$ should be fixed from the experimental data. Due to
lack of sufficient experimental evidence, in many parametrizations,
which are formulated as initial conditions for evolution equations~(\ref{eq:Pom},~\ref{eq:Odd}),
the parameter $\lambda$ remains largely unconstrained (provided $\lambda$
is not very large, so that the identity $\mathcal{O\ll\mathcal{N}}$
remains valid). For this reason, the predictions done with such parametrizations
are given up to global normalization factor. In this work, we explore
two popular parametrizations of the Odderon amplitude, available from~\cite{Benic:2023ybl}
and~~\cite{Hatta:2005as,Yao:2018vcg} described below.
\subsubsection{The odderon parametrization from Ref.~\cite{Benic:2023ybl}}

The parametrization proposed in Ref.~\cite{Benic:2023ybl} incorporates
the transverse geometry of the proton through a Gaussian profile and
introduces a local saturation scale that depends on both the dipole
size and the impact parameter. At the initial scale $Y_{0}$, which
corresponds to $x\approx10^{-2}$, the odderon amplitude is chosen
as 
\begin{eqnarray}
\begin{split}{\cal O}(Y_{0},\,\boldsymbol{r},\boldsymbol{b}) & =\frac{\Lambda_{{\rm JV}}}{8}\left[R_{A}\frac{{\rm d}T_{A}(b)}{{\rm d}b}A^{2/3}\frac{\sigma_{0}}{2}\right]Q^{3}_{S,0}A^{1/2}r^{3}(\hat{\boldsymbol{r}}\cdot\hat{\boldsymbol{b}})\log\left(\frac{1}{r\Lambda_{{\rm QCD}}}+e_{c}{\rm e}\right)\exp\left[-\frac{1}{4}r^{2}Q^{2}_{0,p}\left(\boldsymbol{r},\,\boldsymbol{b}\right)\right]\,\end{split}
\label{eq:oddben}
\end{eqnarray}
where the exponential suppression of large dipoles is controlled by
the local saturation scale $Q^{2}_{0,p}\left(\boldsymbol{r},\,\boldsymbol{b}\right)$,
\begin{align}
Q^{2}_{0,p}\left(\boldsymbol{r},\,\boldsymbol{b}\right) & \equiv A\,T_{p}\left(\boldsymbol{b}\right)\,\frac{\sigma_{0}}{2}\,Q^{2}_{S,0}\log\left(\frac{1}{r\Lambda_{{\rm QCD}}}+e_{c}\,e\right),\,\,T_{p}\left(\boldsymbol{b}\right)=\frac{1}{\pi R^{2}_{p}}\exp\left(-\frac{b^{2}}{R^{2}_{p}}\right),\label{eq:Q0p}\\
 & Q^{2}_{S,0}=0.06~\text{GeV}^{2},\quad e_{c}=18.9,\quad\frac{\sigma_{0}}{2}=16.36~\text{mb}.
\end{align}
the proton radius $R_{p}$ is fixed using the relation $\pi R^{2}_{p}=\sigma_{0}/2=4\pi B_{p}$,
and the value of the constant $\Lambda_{{\rm JV}}$ for the sake of
definiteness was chosen as in the Jeon-Venugopalan model~\cite{Jeon:2005cf,Benic:2023ybl}

\begin{eqnarray}
\lambda=\Lambda_{{\rm JV}}=-\frac{3}{16}\frac{N^{2}_{c}-4}{(N^{2}_{c}-1)^{2}}\frac{Q^{3}_{S,0}A^{1/2}R^{3}_{A}}{\alpha^{3}_{S}A^{2}}\,.\label{eq:lajv}
\end{eqnarray}
In view of the above-mentioned normalization uncertainty, we will
replace $\Lambda_{{\rm JV}}$ with $\lambda\,\Lambda_{{\rm JV}}$
and will treat $\lambda$ as a free parameter whose values can change
in a certain range dictated by consistency with experimental $pp$
and $p\bar{p}$ data.

\subsubsection{The odderon parametrization from Ref.~~\cite{Hatta:2005as,Yao:2018vcg}}

\label{subsec:hatta}

The parametrization proposed in~\cite{Hatta:2005as,Yao:2018vcg}
does not take into account the impact parameter dependence of the
dipole and considers that the odderon amplitude $\mathcal{O}$ has
a nontrivial dependence on the angle between the dipole separation
$\boldsymbol{r}$ and the transverse spin $\boldsymbol{S}$ of the
target. This parametrization is better suited for description of the
observables which include explicitly the dependence on the spin of
the target, such as the gluon Sivers function. Explicitly, the parametrization
is given by

\begin{equation}
{\cal O}\left(Y_{0},\,\boldsymbol{r}\right)=2\kappa\,Q^{3}_{s0}r^{2}\left[\boldsymbol{\mathcal{S}}\times\boldsymbol{r}\right]_{z}\exp\left(-r^{2}Q^{2}_{s0}\right)=\kappa\,Q^{3}_{s0}r^{3}\sin\theta\exp\left(-r^{2}Q^{2}_{s0}\right),\label{OddHatta}
\end{equation}
where $\theta$ is the angle between the vectors $\boldsymbol{\mathcal{S}}$
and $\boldsymbol{r}$, the variable $Q_{s0}\approx1\,{\rm GeV}$ is
the initial saturation scale, and $\kappa$ is a dimensionless parameter
controlling the normalization and the sign of the Odderon amplitude.
While in~\cite{Yao:2018vcg} the value of $\kappa=1/3$ was used
for definiteness, as we discussed earlier, we will assume that it
may be multiplied by an arbitrary constant $\lambda$. The lack of
explicit $b$-dependence in the parametrization~(\ref{OddHatta})
is not relevant for inclusive observables. However, for exclusive
processes the $b$-dependence plays a crucial role and determines
the dependence of the cross-section on the invariant momentum transfer
$t$. For this reason, we introduce a multiplicative impact-parameter
dependent profile $S(b)$ into~(\ref{OddHatta}), namely replace~(\ref{OddHatta})
with 
\begin{equation}
{\cal O}\left(Y_{0},\,\boldsymbol{r},\,\boldsymbol{b}\right)\;=\lambda\,\kappa\,Q^{3}_{s0}r^{3}\left(\hat{\boldsymbol{r}}\cdot\hat{\boldsymbol{b}}\right)\exp\left(-r^{2}Q^{2}_{s0}\right)\;S(b).\label{OddHatta_b}
\end{equation}
For the profile $S(b)$ we analyze different parametrizations of $S(b)$
suggested in~Refs.~\cite{Roa:2023skv,Rezaeian:2013tka,Contreras:2021whz}
in the context of modeling the dipole scattering amplitude and summarized
in Table~\ref{tab:profiles}. The parameter $\lambda$ (its upper
possible value) will be fixed later from the experimental fit of $pp$
and $p\bar{p}$ data. Since in what follows we will consider only
unpolarized observables, in~(\ref{OddHatta_b}) we also slightly
modified the dependence on the orientation of $\boldsymbol{r}$ and
replaced the spin-dependent factor $\left[\boldsymbol{\mathcal{S}}\times\hat{\boldsymbol{r}}\right]_{z}$
with $\left(\hat{\boldsymbol{r}}\cdot\hat{\boldsymbol{b}}\right)$.
We can see that the amplitude~(\ref{OddHatta_b}) satisfies the desired
antisymmetry property $\mathcal{O}\left(-\boldsymbol{r},\,\boldsymbol{b}\right)=-\mathcal{O}\left(\boldsymbol{r},\,\boldsymbol{b}\right)$
of the odderon amplitude.

\begin{table}[ht]
{\renewcommand{\arraystretch}{1.5}\centering {\footnotesize{}}{\footnotesize{}%
\begin{tabular}{|c|c|c|}
\hline 
{\footnotesize Refs} & {\footnotesize Profile function $S(b)$} & {\footnotesize$m$}\tabularnewline
\hline 
{\footnotesize 1.-~~\cite{Roa:2023skv}~~~~} & {\footnotesize$mbK_{1}(mb)$} & {\footnotesize 0.5412 GeV}\tabularnewline
\hline 
{\footnotesize 2.-~~\cite{Contreras:2021whz}~~~~} & {\footnotesize$\text{Exp}(-mb)$} & {\footnotesize 0.840 GeV}\tabularnewline
\hline 
{\footnotesize 3.-~~\cite{Rezaeian:2013tka}~~~~} & {\footnotesize$\text{Exp}(-mb^{2})$} & {\footnotesize 0.070 GeV$^{2}$}\tabularnewline
\hline 
\end{tabular}} \caption{Impact-parameter profile functions $S(b)$ used in this work. In what
follows we will refer to them as $S_{1}(b),$ $S_{2}(b)$ and $S_{3}(b)$,
respectively. The effective parameter $m$ determines suppression
of the odderons at large values of the impact parameter $b$.}
\label{tab:profiles} }
\end{table}

\section{Elastic $pp$ scattering}

\label{sec:amplitude}

As we mentioned in the Introduction, the odderons have been predicted
many years ago, however for a long time any attempts to detect them
experimentally yielded only upper limits, and at present the only
compelling indication in favor of odderons is the mismatch of the
$pp$ and $p\bar{p}$ elastic cross-cross-sections measured by the
TOTEM~\cite{TOTEM:2018psk} and D0~\cite{D0:2012erd} collaborations.
For this reason, it is not surprising that the estimates for the odderon
amplitude vary significantly and include unknown constants. To the
best of out knowledge, most of the previous attempts to describe these
data were based on the effective models that encode $C$-odd and $C$-even
parts of the elastic $pp$ amplitude~\cite{Luna:2024cbq,Khoze:2018kna,Broilo:2020kqg,Gotsman:2020mkd}.
However, the main limitation of those studies is the lack of universality:
its extension to other odderon-mediated processes is not straightforward,
whereas extrapolation to different energies relies on additional assumptions
that should be fixed from experimental data, rather than from the
QCD evolution equations.

\begin{figure}
\includegraphics[totalheight=5cm]{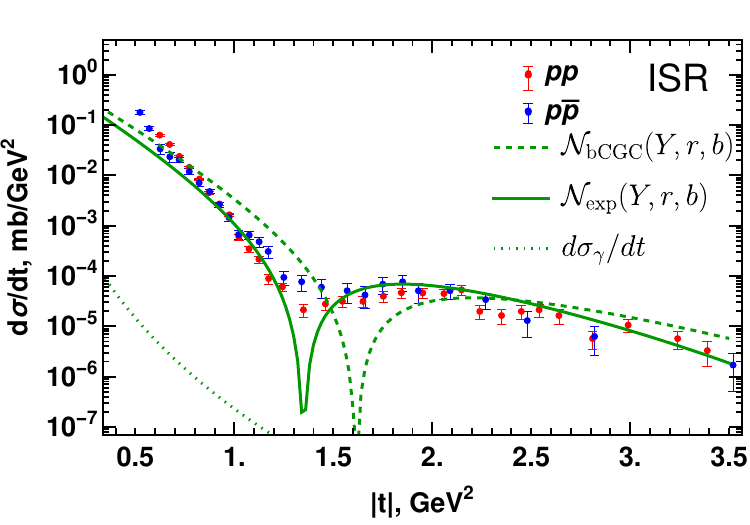}\includegraphics[totalheight=5cm]{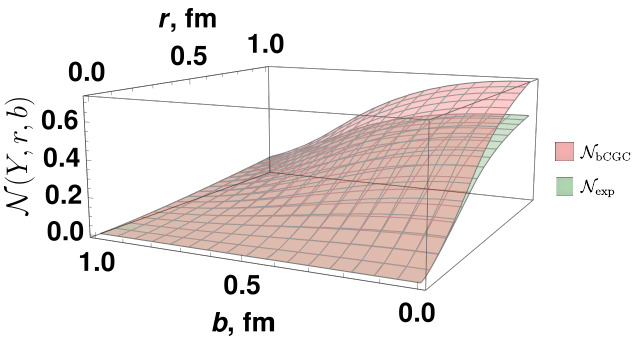}

\caption{(Color online) Left: Contribution of the dominant ($C$-even) part
of the dipole amplitude to the elastic cross-section $d\sigma/dt$.
The dotted curve $d\sigma_{\gamma}/dt$ is added for reference and
corresponds to the photon-only contribution. We use notations $\mathcal{N_{{\rm bCGC}}}$
for the bCGC parametrization of the dipole amplitude from~\cite{Rezaeian:2013tka},
and $\mathcal{N}_{{\rm exp}}=1-\exp\left(-\mathcal{N_{{\rm bCGC}}}\right)$
for the exponentiated form suggested in~\cite{Levin:1,Levin:2} (it
reduces to $\mathcal{N_{{\rm bCGC}}}$ for small-size dipoles that
dominate in DIS kinematics). The evaluation was performed using the
wave functions (\ref{LFWFb}). A qualitatively similar behavior was
observed for the TOTEM-D0 kinematics. Right plot: Comparison of the
two parametrizations $\mathcal{N_{{\rm bCGC}}}$ and $\mathcal{N}_{{\rm exp}}$
in configuration space. The plot illustrates that the parametrizations
nearly coincide for small-size dipoles. See the text for more details.}
\label{fig:elastic}
\end{figure}

In our phenomenological analysis, we largely follow the approach outlined
in~\cite{Hagiwara:2020mqb} and compute rigorously the differential
cross section $d\sigma/dt$ for the elastic proton-proton and proton-antiproton
scattering using the color glass condensate framework. We will consider
the $pp$ and $p\bar{p}$ as a dilute-dense system, treating one of
the hadrons (projectile) as an ensemble of partons (quark-diquark
system), whose scattering in the gluonic field of the other hadron
is described by the pomeron and odderon scattering amplitudes. According
to canonical CGC rules, the interaction of the color singlet 3-quark
system with the target should be described by the correlator
\begin{equation}
S\left(\boldsymbol{x}_{1},\,\boldsymbol{x}_{2},\,\boldsymbol{x}_{3}\right)=\frac{1}{3!}\epsilon_{b_{1}b_{2}b_{3}}\epsilon^{a_{1}a_{2}a_{3}}\left\langle U^{b_{1}}_{a_{1}}\left(\boldsymbol{x}_{1}\right)U^{b_{2}}_{a_{2}}\left(\boldsymbol{x}_{2}\right)U^{b_{3}}_{a_{3}}\left(\boldsymbol{x}_{3}\right)\right\rangle \label{eq:S}
\end{equation}
where the Levi-Civita symbol $\epsilon_{a_{1}a_{2}a_{3}}$ reflects
antisymmetrization with respect to color singlet indices, and the
angular brackets denote averaging over all color sources. The color
averaging leads to a strong suppression of the color charges separated
by distances $\left|\boldsymbol{x}_{i}-\boldsymbol{x}_{j}\right|\gtrsim Q^{-1}_{s}$,
where $Q_{s}$ is the saturation scale (inverse of the color correlation
length). As was discussed in~~\cite{Hatta:2005as,Balitsky:2014mca},
in general this correlator does not reduce to the ``dipolar'' amplitudes
$\mathcal{N},\,\mathcal{O}$, but may also include additional quadrupolar
contributions, which mix up non-trivially in the process of evolution.
Since the early days of QCD, it has been understood that for many
high-energy processes, the hard interaction involves a single active
quark within the proton, while the remaining two act as a joint source
of the color field with nearly coinciding coordinates. In view of
the $SU(3)$ identity $\epsilon_{a_{1}a_{2}a_{3}}U^{b_{1}}_{a_{1}}\left(\boldsymbol{x}\right)U^{b_{2}}_{a_{2}}\left(\boldsymbol{x}\right)\approx\epsilon^{b_{1}b_{2}b_{3}}\left[U^{\dagger}\left(\boldsymbol{x}\right)\right]^{a_{3}}_{b_{3}}$,
a pair of overlapping quarks in the three-quark ensemble is effectively
replaced by a single diquark, so the proton is treated as a bound
quark-diquark state~\cite{GellMann:1964ewy,Ida:1966ev,Lichtenberg:1967zz}.
For such configurations, the correlator~(\ref{eq:S}) effectively
reduces to a combination of the ``dipolar'' amplitudes $\mathcal{N},\,\mathcal{O}$,
and the cross-sections of the elastic $pp$ and $p\bar{p}$ scattering
may be represented as~~\cite{Levin:1,Levin:2,Levin:3}
\begin{equation}
\frac{d\sigma^{(\pm)}}{dt}=\frac{1}{16\pi}\left|\mathcal{A}^{(\pm)}\left(s,\,t\right)\right|^{2}\label{eq:XSec}
\end{equation}
where the amplitude is given by
\begin{equation}
\mathcal{A}^{(\pm)}(t)\,=\,i\int d^{2}\boldsymbol{r}\,d^{2}\boldsymbol{b}\,dx\,\left|\psi_{2}(x,\,\boldsymbol{r})\right|^{2}\,{\rm e}^{-i\left(\boldsymbol{b}\,+\,x\,\boldsymbol{r}\right)\cdot\boldsymbol{\Delta}}\left[\mathcal{N}(Y,\,\boldsymbol{r},\,\boldsymbol{b})\mp i\,{\cal O}(Y,\,\boldsymbol{r},\,\boldsymbol{b})\right]\pm\mathcal{A}_{\gamma}(t),\label{Amp1}
\end{equation}
and the choice of upper or lower sign in $\pm,\mp$ corresponds to
$pp$ or $p\bar{p}$ , respectively. The variable $\del$ in~(\ref{Amp1})
is the transverse momentum transfer between the two hadrons (it is
related to the Mandelstam variable $t$ by $t=-\del^{2}$).
The last term in~(\ref{Amp1}) corresponds to the electromagnetic
contribution with photon exchange in the $t$-channel, which potentially
can also contribute to odderon-mediated processes, as was discussed
long ago in~\cite{Buttimore:1978ry}. Its evaluations is straightforward
in helicity basis using the light-cone rules from~\cite{Lepage:1980fj,Brodsky:1997de}.
As was demonstrated in~\cite{Hagiwara:2020mqb}, the helicity flip
of each nucleon decreases the amplitude $\mathcal{A}_{\gamma}(t)$
by the factor $\sim\,\sqrt{|t|/4M^{2}}\,F_{2}\left(t\right)/F_{1}\left(t\right)$,
which is small in the kinematics of interest~\cite{JeffersonLabHallA:1999epl,PDG}.
For the sake of simplicity, in what follows we will focus on the helicity
non-flip part, which is given explicitly in the single-photon approximation
by~\cite{Hagiwara:2020mqb}~\footnote{Since we use nonrelativistic normalization of the bra- and ket states,
the amplitude differs from~\cite{Hagiwara:2020mqb} by factor $2/s$,
where the Mandelstam variable $s$ is the invariant energy of the
collision,}

\begin{equation}
\mathcal{A}_{\gamma}(t)=4\pi\alpha_{\text{em}}\frac{4}{t}F^{2}_{1}(t),\label{photonHatta}
\end{equation}
where $\alpha_{\text{em}}$ is the fine structure constant, and $F_{1}(t)$
stands for the Dirac form factor of the proton. Since we do not consider
very small $|t|$ in our analysis, we do not take into account the
so called Coulomb phase factor $e^{i\alpha_{{\rm em}}\phi_{c}}$ which
was evaluated in~\cite{Bethe,WestYennie,Buttimore:1978ry} and takes
into account certain higher order corrections. In order to demonstrate
the validity of the quark-diquark approximation, in the Figure~\ref{fig:elastic}
we have shown the corresponding contributions of the dominant ($C$-even)
amplitude $\mathcal{N}$ to the cross-sections $d\sigma/dt$. We may
observe that this approach provides a reasonable qualitative description
of the data. The diffractive minimum (dip) occurs due to cancellation of contributions of different kinematic regions in the integral~(\ref{Amp1}) and
is very sensitive to the $b$-dependence implemented in dipole amplitude
for large-size dipoles. The region near the dip obtains the dominant
contribution from various subleading mechanisms, and can have enhanced
sensitivity to various $C$-odd contributions. We also need to mention
that the above-mentioned uncertainty in $b$-dependence implemented
in dipole parametrization for large dipoles has only minimal influence
for analysis of the $C$-odd sector because the last two terms in~(\ref{eq:Odd})
in the saturated regime give a vanishingly small contribution. This
happens because the $C$-even part $\mathcal{N}$ is approximately
constant in the saturated regime (so it can be taken out of the integrand),
and the remaining integral vanishes in view of the antisymmetry property
of the odderon amplitude, $\mathcal{O}\left(Y,\,-\boldsymbol{r},\,\boldsymbol{b}\right)=-\mathcal{O}\left(Y,\,\boldsymbol{r},\,\boldsymbol{b}\right)$. 

\subsection{Observable for extracting the $C$-odd contribution}

\label{subsec:Isolating_Odd} In the elastic cross-section~(\ref{eq:XSec},~\ref{Amp1})
the contribution of the odderon is commingled with that of pomeron
and photon, which complicates the analysis. At present the strongest
evidence in favor of odderons is the mismatch of TOTEM~\cite{TOTEM:2018psk}
and D0~\cite{D0:2012erd,D0:2020tig} data near the so-called diffractive
minimum of the elastic cross section, at relatively large $|t|\in\left(0.5-1\right){\rm GeV}^{2}$.
However, even in this kinematics the odderons still remain a small
correction and reveal themselves only from mismatch of $pp$ and $\bar{p}p$
data. We will not consider in this manuscript the contribution of
odderons to the ratio of the real to imaginary parts of the forward
scattering amplitude, because the latter may be strongly influenced
by other mechanisms (e.g. contributions of the so-called secondary
odderons~\cite{Gotsman:2018buo}). Since in~~(\ref{Amp1}) the
odderon and photon contribute with the same sign, for simplicity we
will split the amplitude~(\ref{Amp1}) into the $C$-odd and $C$-even
parts, 
\begin{equation}
\mathcal{A}_{\mathbb{O}}(t)\,=\,-\int d^{2}\boldsymbol{r}\,d^{2}\boldsymbol{b}\,dx\,\left|\psi_{2}(x,\,\boldsymbol{r})\right|^{2}\,{\rm e}^{-i\left(\boldsymbol{b}\,+\,x\,\boldsymbol{r}\right)\cdot\boldsymbol{\Delta}}{\cal O}(Y,\,\boldsymbol{r},\,\boldsymbol{b})+\mathcal{A}_{\gamma}(t),\label{Amp1-1}
\end{equation}

\begin{equation}
\mathcal{A}_{\mathbb{N}}(t)\,=\,\left(\rho+i\right)\int d^{2}\boldsymbol{r}\,d^{2}\boldsymbol{b}\,dx\,\left|\psi_{2}(x,\,\boldsymbol{r})\right|^{2}\,{\rm e}^{-i\left(\boldsymbol{b}\,+\,x\,\boldsymbol{r}\right)\cdot\boldsymbol{\Delta}}\mathcal{N}_{{\rm DIS}}(Y,\,\boldsymbol{r},\,\boldsymbol{b}),\label{Amp1-2}
\end{equation}
tacitly assuming that the contribution of the ``pure'' odderon can
be obtained from $\mathcal{A}_{\mathbb{O}}$ subtracting the well-known
photon amplitude. In the $C$-even contribution we replaced the dipole
amplitude $\mathcal{N}$ with its imaginary part $\mathcal{N}_{{\rm DIS}}(Y,\,\boldsymbol{r},\,\boldsymbol{b})$
measurable in Deep Inelastic Scattering, multiplied by additional
prefactor
\begin{equation}
\mathcal{N}(Y,\,\boldsymbol{r},\,\boldsymbol{b})=\left(\rho+i\right)\mathcal{N}_{{\rm DIS}}(Y,\,\boldsymbol{r},\,\boldsymbol{b}),\quad\rho\equiv\frac{{\rm Re\,\mathcal{A}}}{{\rm Im}\,\mathcal{A}}.\label{eq:N}
\end{equation}
The elastic cross-section~(\ref{eq:XSec}) in terms of these notations
can be rewritten as 
\begin{equation}
\frac{d\sigma^{(\pm)}}{dt}=\frac{1}{16\pi}\left[\left|\mathcal{A}_{\mathbb{N}}\right|^{2}+\left|\mathcal{A}_{\mathbb{O}}\right|^{2}\pm2\,\mathrm{Re}\left(\mathcal{A}^{*}_{\mathbb{N}}\mathcal{A}_{\mathbb{O}}\right)\right].\label{eq:XSecSimplified}
\end{equation}
If we take into account the smallness of the amplitude $\mathcal{A}_{\mathbb{O}}$
and disregard the term $\left|\mathcal{A}_{\mathbb{O}}\right|^{2}$
in~(\ref{eq:XSecSimplified}), we may obtain for the sum and the
difference of the $pp$ and $p\bar{p}$ cross-sections

\begin{equation}
\frac{d\sigma^{(+)}}{dt}+\frac{d\sigma^{(-)}}{dt}\approx\frac{1}{8\pi}\,\left|\mathcal{A}_{\mathbb{N}}\right|^{2},\qquad\frac{d\sigma^{(+)}}{dt}-\frac{d\sigma^{(-)}}{dt}=\frac{1}{4\pi}\mathrm{Re}\left(\mathcal{A}^{*}_{\mathbb{N}}\mathcal{A}_{\mathbb{O}}\right).\label{eq:diff}
\end{equation}
Using symmetries of $\mathcal{O}$, it is possible to show that the
amplitude $\,\mathcal{A}_{\mathbb{O}}$ is a real-valued function,
so we may rewrite~(\ref{eq:diff}) as 
\begin{equation}
\Sigma=\frac{1}{\sqrt{2\pi}}\frac{\rho}{\sqrt{1+\rho^{2}}}\mathcal{\mathcal{A}_{\mathbb{O}}}=\frac{d\sigma^{(pp)}/dt-d\sigma^{(p\bar{p})}/dt}{\sqrt{d\sigma^{(pp)}/dt+d\sigma^{(p\bar{p})}/dt}},\label{eq:Razr}
\end{equation}
which allows to relate directly the effective odderon amplitude $\mathcal{\mathcal{A}_{\mathbb{O}}}$
with physically measurable cross-sections. Since this observable does
not include the contribution of pomerons, we believe that it is better
suited for phenomenological studies since it allows to single out
a feeble odderon signal. We may observe that the coefficient in front
of $\mathcal{\mathcal{A}_{\mathbb{O}}}$ depends on the ratio $\rho$
of real to imaginary parts of the dipole amplitudes defined in~(\ref{eq:N}).
At very high energies it is expected that the amplitude is dominated
by the imaginary part, so the parameter $\rho\ll1$, thus leading
to a strong suppression of the odderon signal. While it is possible
to estimate the ratio of the real and imaginary parts $\rho$ using
dispersion relations~\cite{Bronzan:1974jh,Gotsman:1992ui}, this
would require additional model assumptions and introduce dependence
on the parametrization of the dipole amplitude~$\mathcal{N}_{{\rm DIS}}$.
For this reason, in what follows we will use the experimentally measured
value $\rho\approx0.1$~\cite{ATLAS:2022mgx,Amos}.

\section{Comparison of phenomenological models with experimental data}

\label{subsec:data}The main goal of this phenomenological analysis
is to compare the available phenomenological models with experimental
data and find upper limits for the odderon amplitude. For this reason,
we will focus on the mismatch of $pp$ and $p\bar{p}$ cross-section
observed at ISR~\cite{Breakstone:1985}, and a similar mismatch of
D0 and TOTEM experimental data reported in~\cite{Martynov:2017zjz,D0:2020tig,D0:2012erd,TOTEM:2018psk}.
For the TOTEM-D0 odderon measurements, we extracted the numerical
values of the extrapolated $pp$ and measured $p\bar{p}$ cross-sections
directly from Figure~4 of Ref.~\cite{D0:2020tig}. This dataset consists
of sixteen points in total: eight experimental points corresponding
to the D0 $p\bar{p}$ measurement and eight additional points corresponding
to the extrapolated $pp$ cross section from TOTEM. For each data
point, we included both the central value and the experimental uncertainty.
We need to mention that at present the TOTEM data have been questioned
by ATLAS collaboration, and a comparison of their data with D0 suggests
a significantly smaller signal (see~\cite{Csorgo:2024dvr,ATLAS:2022mgx}
for details). While the latter is important for the global analysis
and estimate of statistical significance, in our case this set simply
decreases the upper values of the odderon signal. For the ISR, we
used the experimental points in the range $0.6<|t|<2$ ${\rm GeV^{2}}$
(twenty points in total) because in this range the $pp$ and $p\bar{p}$
cross-sections are measured in the same narrow bins in $|t|$. For
larger $|t|$, the bins used for measurement of $pp$ and $p\bar{p}$
cross-sections do not coincide, and for this reason their comparison
would involve the additional model assumptions that can be detrimental
for studies of the feeble odderon signal. The inclusion of the ISR
data requires due caution. The data from~\cite{Amos} do not present
interest for odderon searches, since in the small-$t$ kinematics
the photon exchange contribution $\mathcal{A}_{\gamma}(t)$ is strongly
enhanced and constitutes the dominant mechanism that explains the
difference of $pp$ and $p\bar{p}$ cross-sections. The data from~\cite{Breakstone:1985}
present more interest, since their kinematic domain partially overlaps
with that of TOTEM-D0. While these data have large systematic uncertainty
(20-30 per cent), the latter is largely due to the global normalization
(scale) uncertainty, and can be directly related to the (systematic)
normalization uncertainty of the odderon signal~(\ref{eq:Razr}).
In what follows we sum the systematic and statistic errors in quadrature.
Despite of the large \emph{relative} error, the ISR data impose useful
upper limits for the magnitude of the odderon amplitude, especially
in the kinematics of large $|t|>1$ GeV$^{2}$, where there is no
other experimental constraints.

\begin{figure}
\begin{tabular}{cc}
\includegraphics[width=17cm]{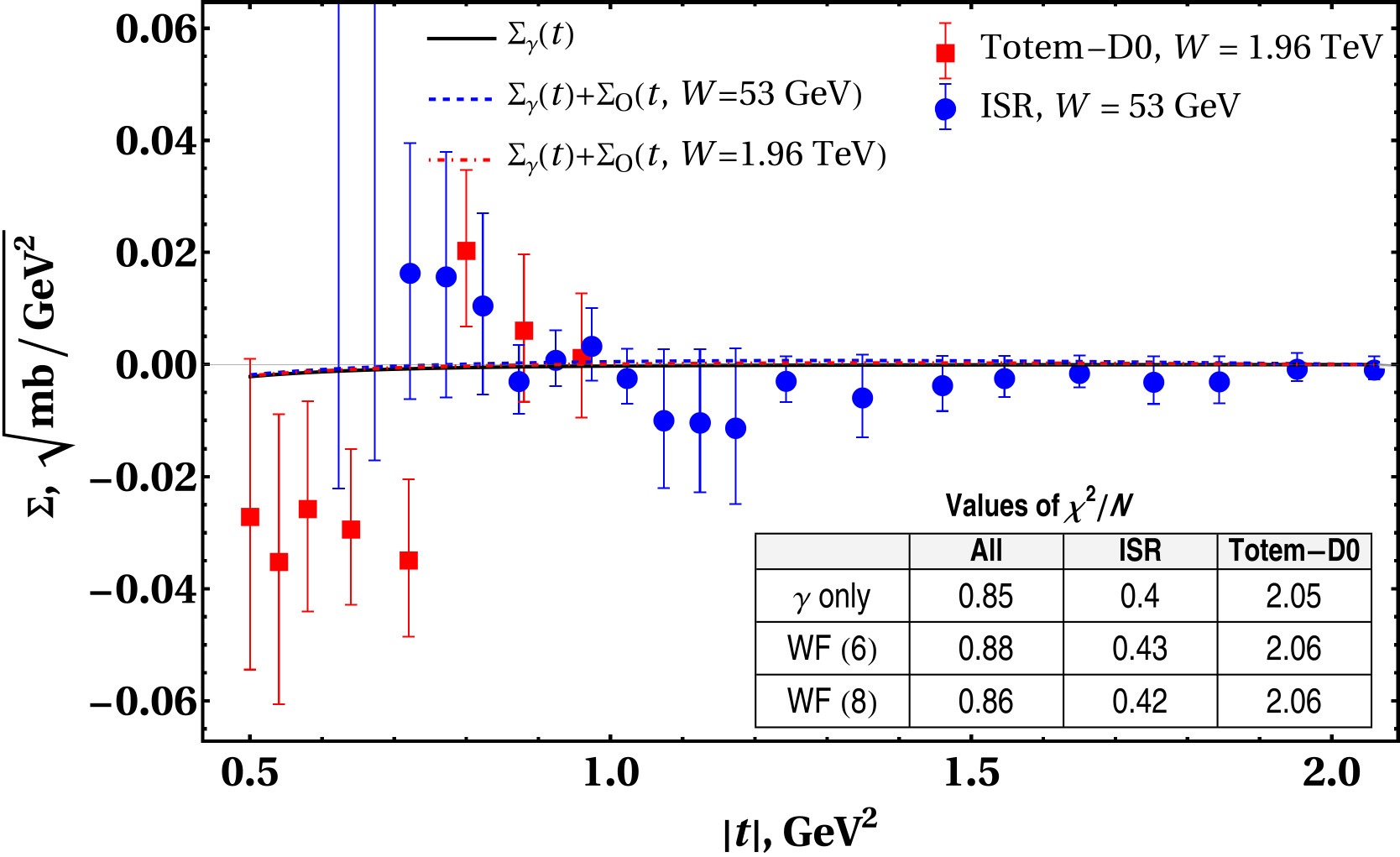} & \tabularnewline
\end{tabular}

\caption{(Color online) The predictions for the odderon-mediated observable
$\Sigma$ with parametrization~(\ref{eq:oddben}) and the wave functions~(\ref{eq:dIgauss},\ref{LFWFb}),
respectively (all curves nearly coincide with the photon-only contribution
$\Sigma_{\gamma}$). The points with error bars (\textquotedblleft experimental
data\textquotedblright{} for the observable $\Sigma$) were derived
from ISR~\cite{Breakstone:1985} and D0-TOTEM experimental data in~\cite{Martynov:2017zjz,D0:2020tig,D0:2012erd,TOTEM:2018psk}
as explained in the text. For ISR, the systematic and statistic errors
were added in quadrature. The solid black curve $\Sigma_{\gamma}(t)$
corresponds to the energy-independent contribution of the $t$-channel
photon exchange, whereas $\Sigma_{\mathbb{O}}(t)$ is the energy-dependent
contribution of the odderon. The quoted values of $\chi^{2}/N$ in
the column \textquotedblleft All\textquotedblright{} of the table correspond
to the global analysis (TOTEM-D0 and ISR), whereas columns \textquotedblleft ISR\textquotedblright{}
and \textquotedblleft Totem-D0\textquotedblright{} take into account
only the data from these collaborations. The first row in the table
gives values for the photon-only contribution, whereas the last two
rows take into account odderons using the wave functions~(\ref{eq:dIgauss})
and (\ref{LFWFb}), respectively.}
\label{fig:odderon_Benic-1}
\end{figure}

As we explained earlier, for each pair of experimental points ($d\sigma_{pp}/dt,\,d\sigma_{p\bar{p}}/dt$)
we construct the observable~(\ref{eq:Razr}) and compare it with
theoretical predictions obtained using different parametrizations
of the odderon amplitudes and wave functions. The relative experimental
error of the derived observable~(\ref{eq:Razr}) increased drastically
due subtraction of very close numbers, in agreement with well-known
results from elementary statistics. From the Figure~\ref{fig:odderon_Benic-1}
we may observe that TOTEM-D0 and ISR data are marginally consistent
with each other in the kinematics $|t|<0.7\,{\rm GeV^{2}}$: it may
be nearly impossible to find a parametrization of the odderon amplitude
$\mathcal{O}(\boldsymbol{r},\,\boldsymbol{b})$ that satisfies the
BK equation and fits simultaneously a homogeneous $t$-dependence
at ISR, and node in the same variable in the kinematics of D0. In
the same Figure~\ref{fig:odderon_Benic-1} we also have shown the
results for the variable $\Sigma$ found with parametrization~(\ref{eq:oddben})
of the odderon amplitude and different wave functions~(\ref{eq:dIgauss},~\ref{LFWFb}).
We solved the BK evolution equations~(\ref{eq:Pom},~\ref{eq:Odd})
numerically, using a three-step third order Adams-Bashforth method,
as described in~\cite{Benic:2023ybl}. For each impact parameter
$b$, we used a lattice with $N_{r}\times N_{\varphi_{r}}=400\times100$
nodes (with logarithmic step in $r$-direction). Since the initial
scale of the parametrization (\ref{OddHatta_b}) was not defined in
the paper, we assumed that evolution is done from the initial scale
$Y_{0}=\ln\left(1/x_{0}\right)$, where $x_{0}=10^{-3}$, and up to
a scale $Y_{{\rm fin}}=\ln\left(s/m^{2}_{N}\right),$ where $\sqrt{s}$
corresponds to 53 GeV for ISR and $1.96\,{\rm TeV}$ for D0. Since
the right-hand side of~ (\ref{eq:Pom},~\ref{eq:Odd}) does not
depend explicitly on $Y$, the effects of evolution depend only on
the difference of rapidities (and initial conditions). In view of
this invariance, the effects of evolution remain invariant under any
redefinition (rapidity shift) $Y\to Y+{\rm const}$. For the dipole
amplitude $\mathcal{N}_{{\rm DIS}}\left(\boldsymbol{r},\,\boldsymbol{b}\right)$
at the initial scale we assume the b-CGC parametrization~\cite{Rezaeian:2013tka},
which yields the constants $\alpha\approx0.15$ and $\rho\approx0.25$.
We found that the evolution suppresses the odderon signal, in agreement
with earlier findings of~\cite{Yao:2018vcg,Hagiwara:2020mqb}. We
may see that the choice of the wave function has only mild influence
on the shape of the $t$-dependence, and mostly affects the magnitude
of the odderon-mediated contribution. This happens because the wave
functions do not depend directly on the impact parameter $b$ (Fourier
conjugate of the variable $\Delta_{\perp}\approx\sqrt{|t|}$), and
the average dipole size is determined by the shape of the odderon
amplitude rather than by the wave function alone. In both cases the
goodness-of-fit metric $\chi^{2}/N$, is close to one, which signals
that both parametrizations describe the experimental data reasonably
well, which might be partially due to large error bars of the experimental
data. For comparison, for the photon-only contribution ($\Sigma_{\gamma}$)
we obtained $\chi^{2}/N=0.85$. If we omit the ISR data with large
error bars and consider only TOTEM-D0 data, then the $\chi^{2}/N$
increases up to $\approx$ 2.05 ($p$-value 0.02), in agreement with~\cite{Cui:2022dcm}.
This signals tension of the photon-only model, and suggests that the
statistical significance of the odderon is $\approx$ 2.3$\sigma$.

\begin{figure}
\begin{tabular}{cc}
\textasciiacute \includegraphics[width=8.4cm]{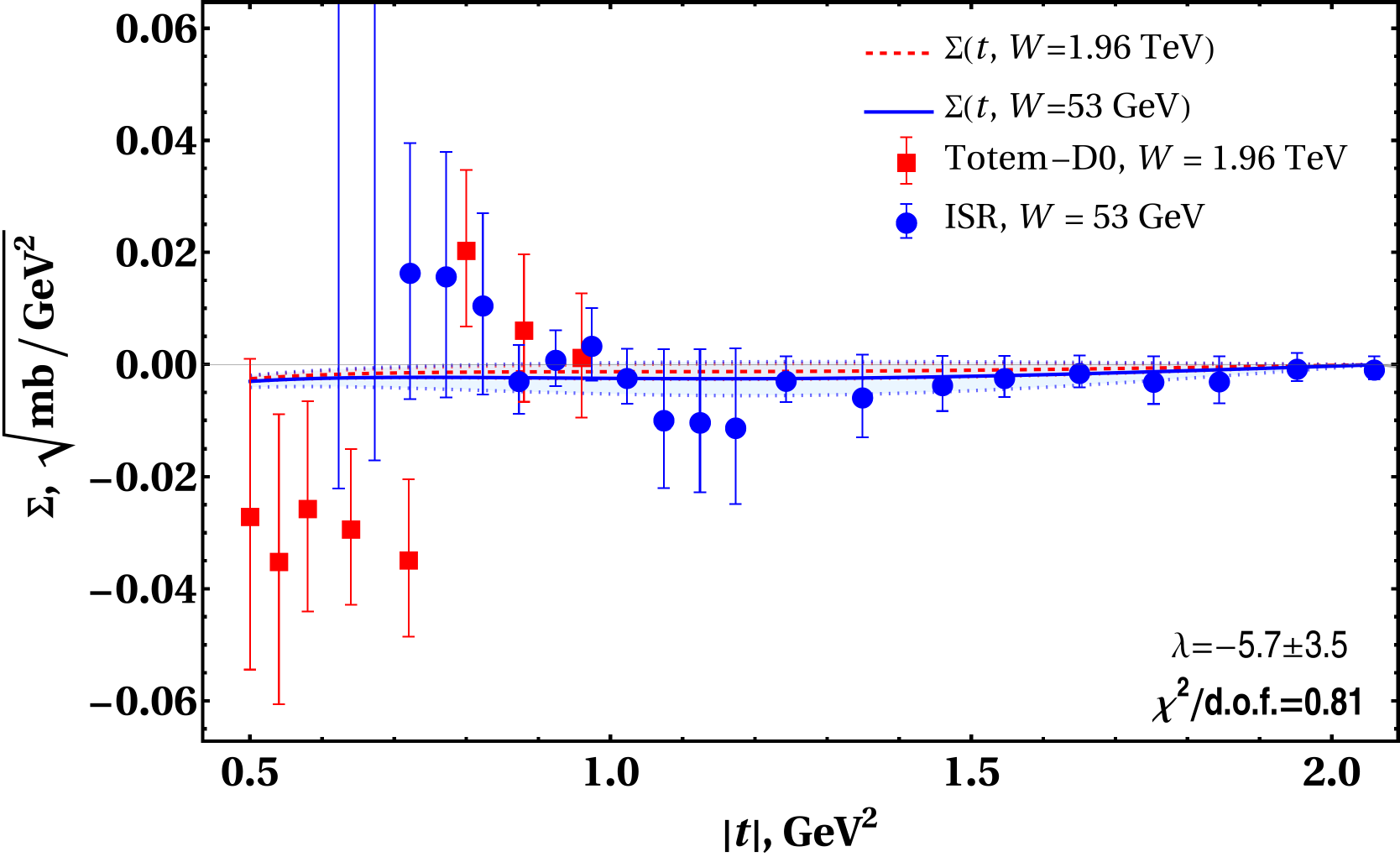} & \includegraphics[width=8.4cm]{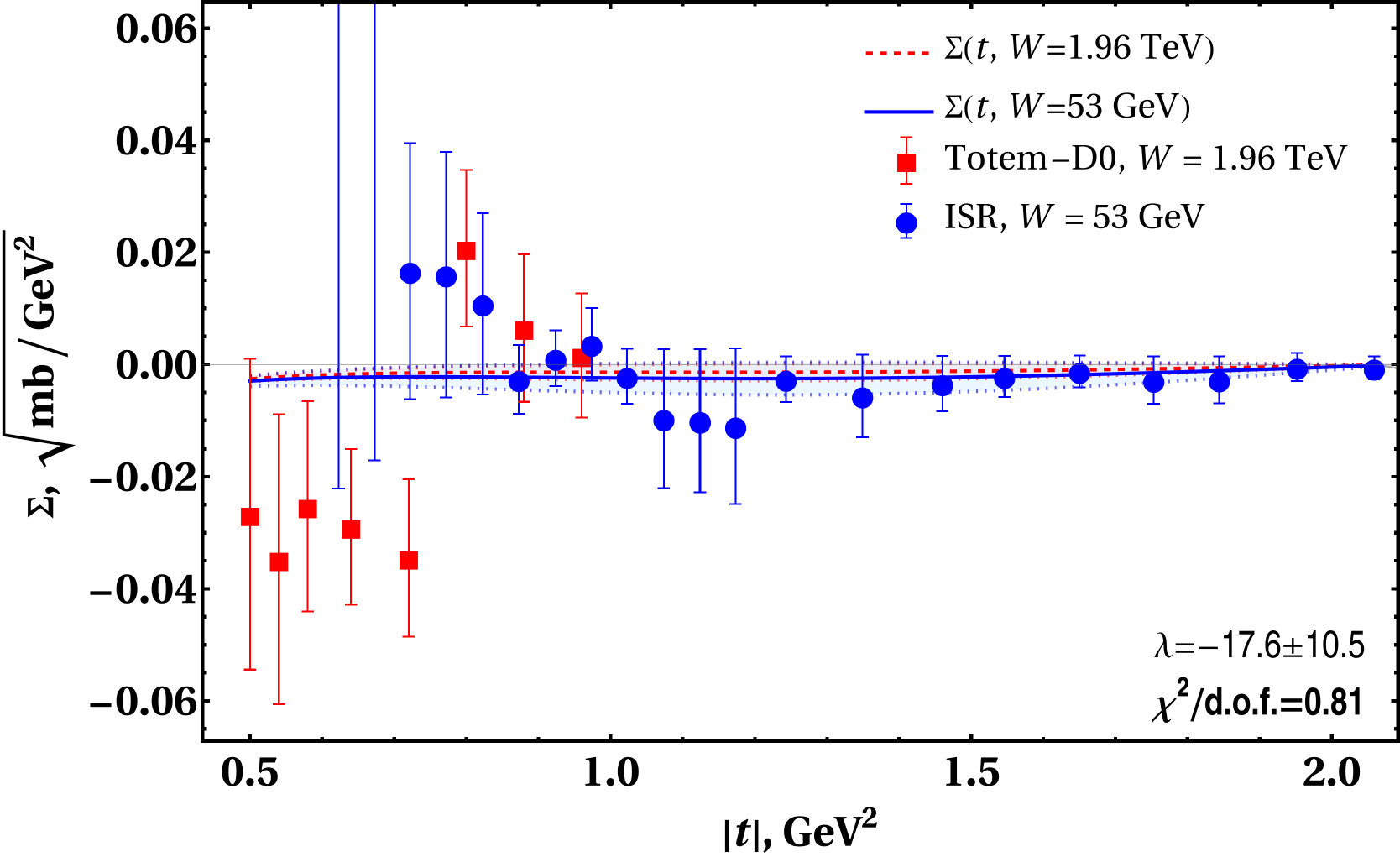}\tabularnewline
\includegraphics[width=8.4cm]{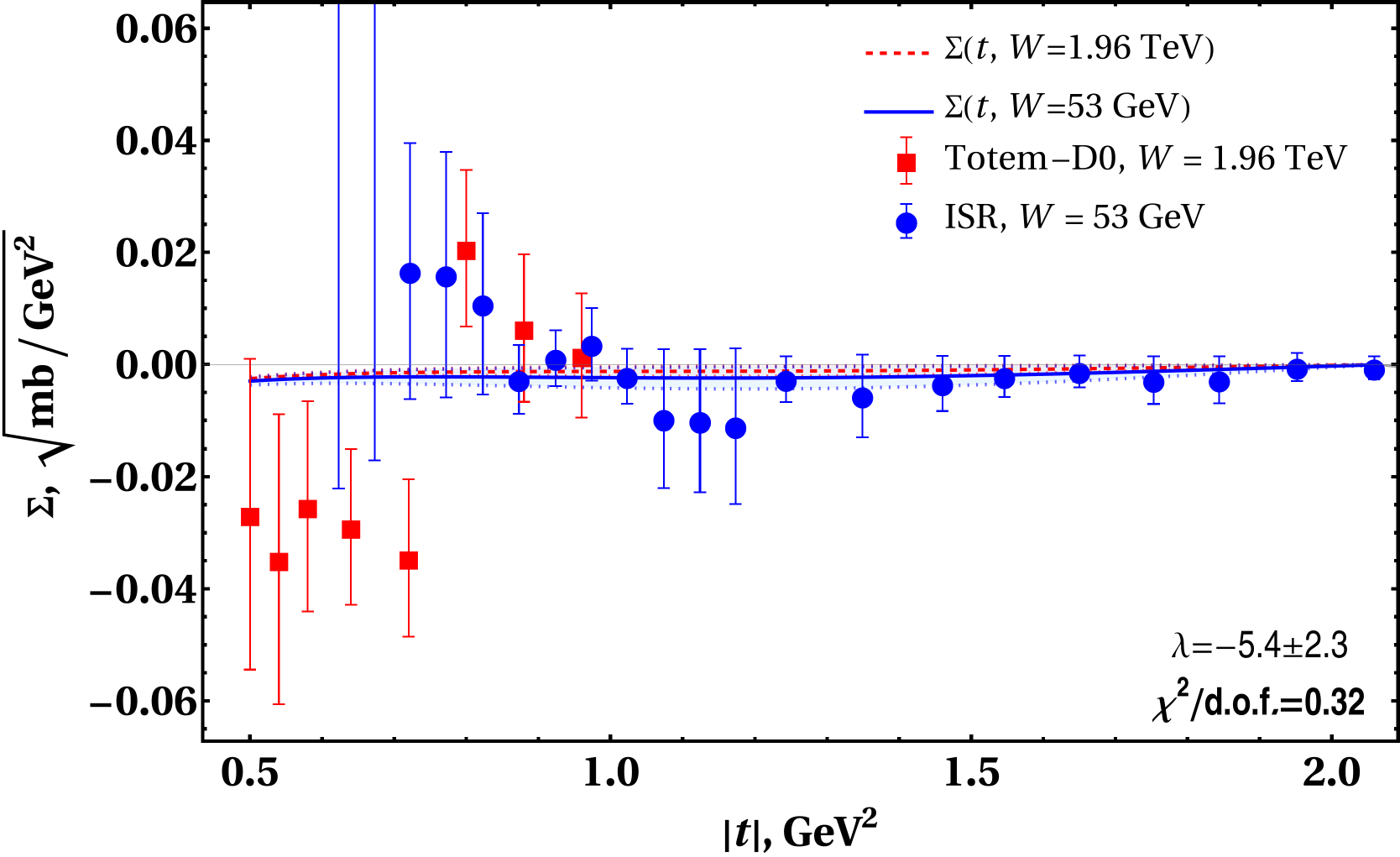} & \includegraphics[width=8.4cm]{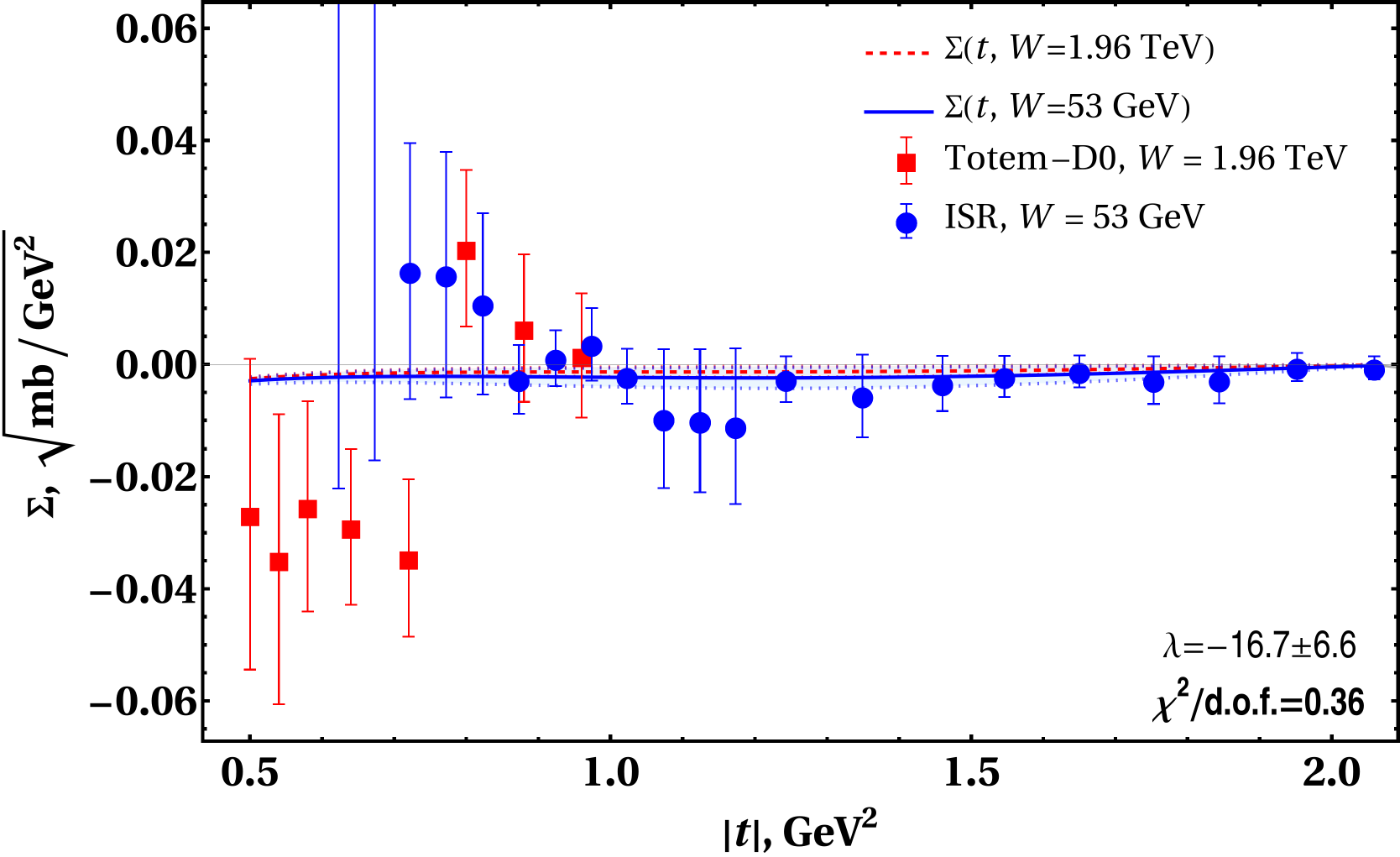}\tabularnewline
\includegraphics[width=8.4cm]{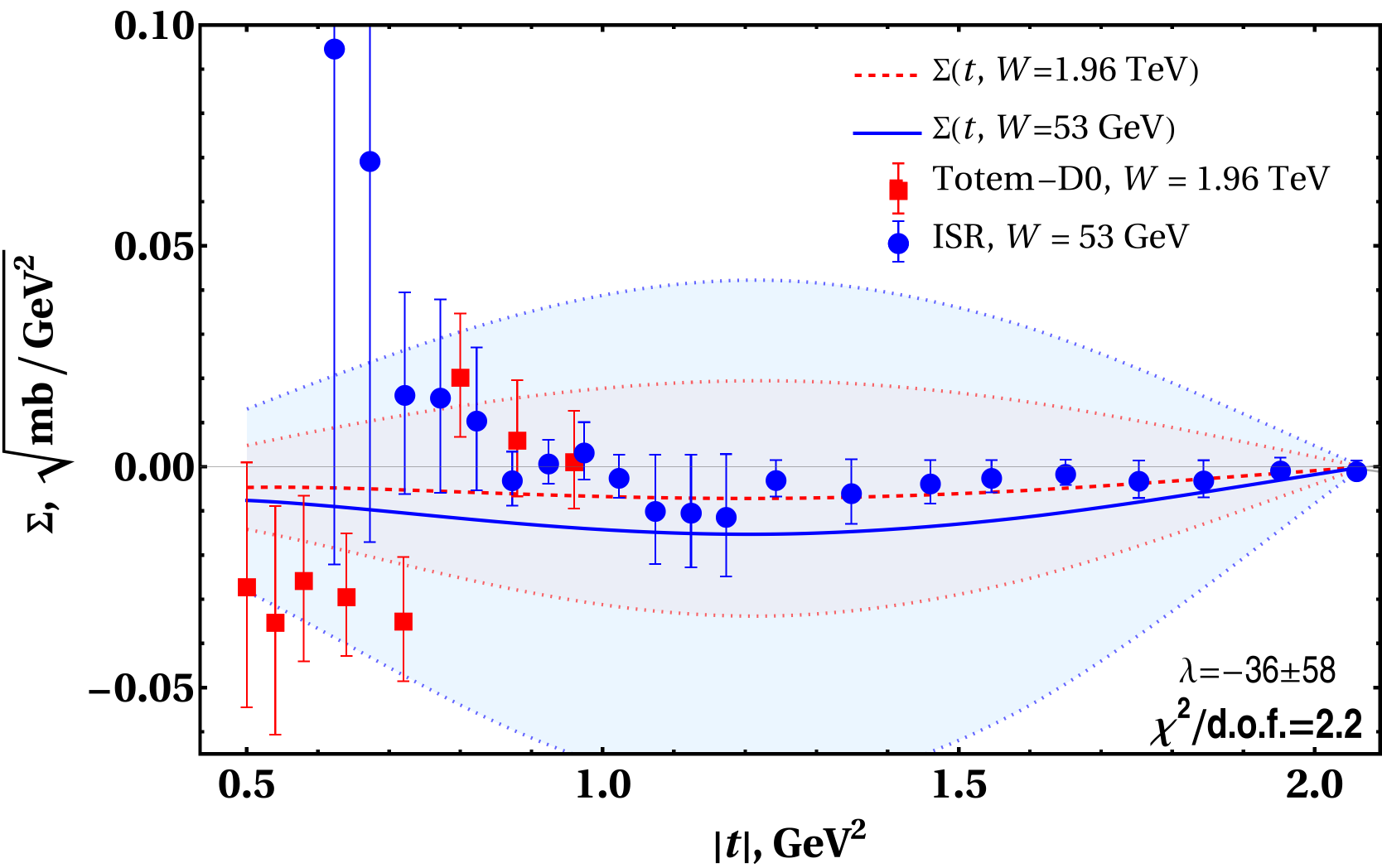} & \includegraphics[width=8.4cm]{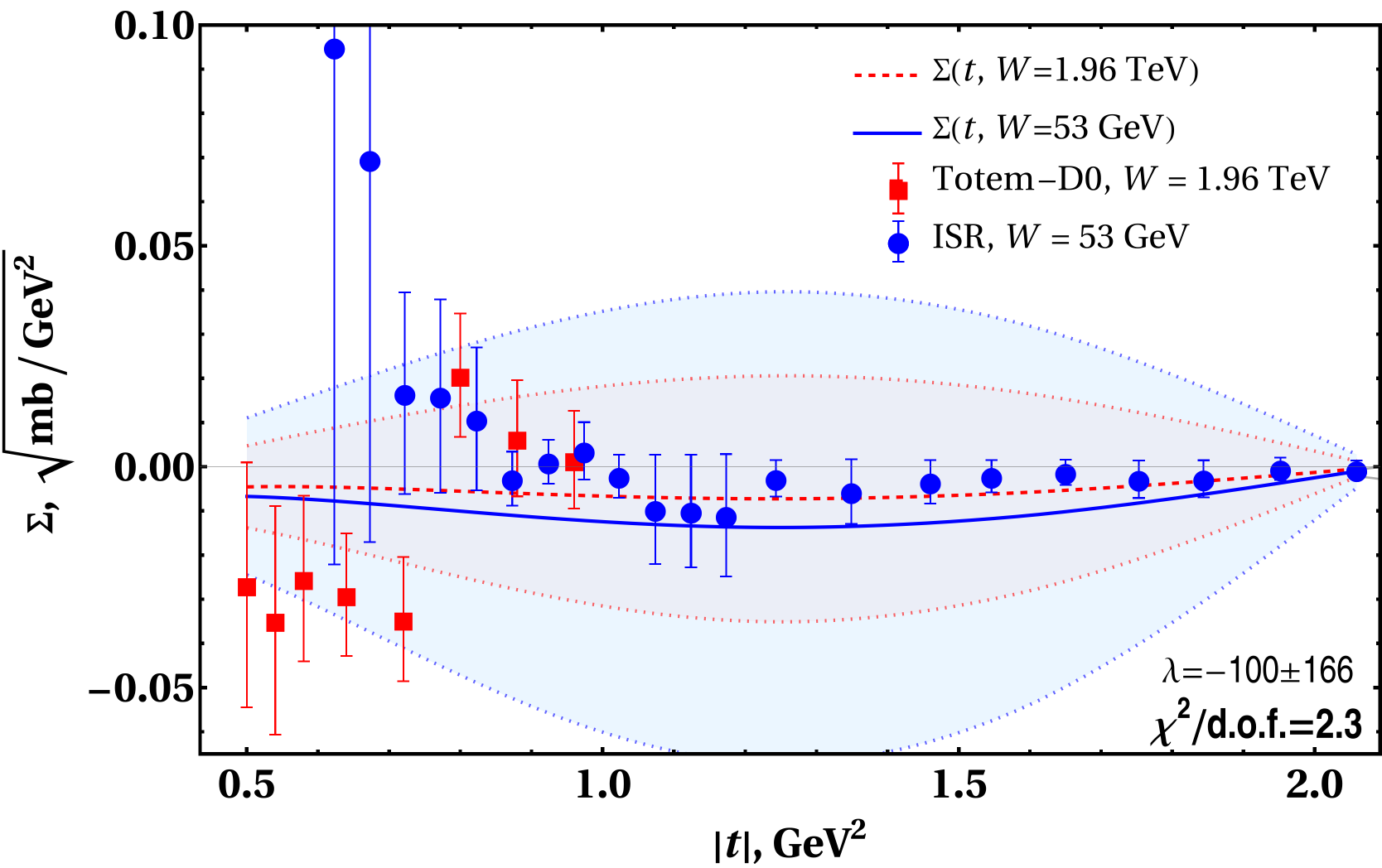}\tabularnewline
\end{tabular}

\caption{(Color online) The predictions for the odderon-mediated observable
$\Sigma$ with parametrization~(\ref{eq:oddben}). The upper row
corresponds to the global fit, the central row corresponds to the
fit of ISR data only, and the last row includes only TOTEM-D0 data
into the fit (see text for details). The left and right columns correspond
to the wave functions~(\ref{eq:dIgauss},\ref{LFWFb}), respectively.
The points with errorbars (\textquotedblleft experimental data\textquotedblright{}
for the observable $\Sigma$) were derived from ISR~\cite{Breakstone:1985}
and D0-TOTEM experimental data in~\cite{Martynov:2017zjz,D0:2020tig,D0:2012erd,TOTEM:2018psk}.
The values of $\lambda$ (together with errors) and the corresponding
$\chi^{2}/{\rm d.o.f.}$ are shown in the lower right corner. The
quoted values of $\chi^{2}/{\rm d.o.f.}$ correspond only to the points
that were taken into account during the fit and should not be confused
with a global goodness-of-fit metrics (see the text for more details).
The values of $\chi^{2}/{\rm d.o.f.}$ shown in the last two plots
exceed $\chi^{2}/N$ shown in the last column of the inline table
in the Figure~\ref{fig:odderon_Benic-1} only due to the difference
of divisors; the corresponding $\chi^{2}/N=1.92$ for TOTEM-D0 fit.
The colored bands show the uncertainty of the fit (95\% confidence
level).}
\label{fig:odderon_Benic}
\end{figure}

We also tried to modify the parametrization~(\ref{eq:oddben}), adding
a free parameter $\lambda$ (normalization constant) in front of it,
as explained in~(\ref{eq:lambda}), and fixing its value from the
global minimization of~$\chi^{2}/{\rm d.o.f.}$ In the Figure~\ref{fig:odderon_Benic}
we have shown the results of the fit, together with obtained value
of $\lambda$ and goodness-of-fit metric $\chi^{2}/{\rm d.o.f.}$,
for different choices of the wave functions and subsets of experimental
data. As we mentioned earlier, the choice of the wave function mostly
affects the magnitude of the odderon-mediated contribution, which
can be absorbed by redefinition of the parameter $\lambda$ that minimizes
the $\chi^{2}/{\rm d.o.f.}$ For this reason, the plots in the left
and the right columns have similar shapes. In the first row of the
Figure~\ref{fig:odderon_Benic} we have shown the results of the
global fit that treats the ISR and TOTEM-D0 data on equal footing.We
may see that minimization does not decrease significantly the values
of $\chi^{2}/{\rm d.o.f.}$ and sets very loose constraints on possible
values of the normalization $\lambda$, although indicates that the
data may be described better with negative and relatively large values
of $\lambda$. Partially such large values are due to the fact that
in the difference of the $pp$ and $p\bar{p}$ cross-sections~(\ref{eq:Razr})
the odderon amplitude is multiplied by the small parameter $\rho$.
Despite of large errors, the extracted values of the parameter $\lambda$
are in agreement with each other, which corroborates consistency of
our approach and robustness of our results to the choice of the proton
wave function. In order to understand better which of the two datasets
constrains stronger the odderon amplitude, we also performed similar
fits using separately ISR or TOTEM-D0 data, and present the results
in the second and the third rows of the Figure~\ref{fig:odderon_Benic},
respectively. The fit of ISR data yields approximately the same confidence
interval for $\lambda$, although the values of $\chi^{2}/{\rm d.o.f.}$
are smaller than in the global fit. At the same time, the fit of the
TOTEM-D0 data alone (see the last row) provides a significantly worse
constraints on the parameter $\lambda$ and overall goodness-of-fit
metrics. Such large values of $\lambda$ (of order $\lambda\sim100$)
clearly would lead to unphysically large cross-section for the exclusive
$\pi^{0},\,\eta_{c},\chi_{c}$ production~\footnote{For our estimates of the cross-sections for the exclusive production
of $\pi^{0},\,\eta_{c},\chi_{c}$, we took predictions from~\cite{Benic:2023ybl,Benic:2024pqe}
and scaled them using the factor $\lambda^{2}$. The pion size (r.m.s.
radius) is approximately larger than that of $\eta_{c},\chi_{c}$,
for this reason for our rough estimates we used the wave function
of $\eta_{c}$ with properly adjusted radius and normalization.} in contradiction to experimental upper bounds at HERA~\cite{H1:2002ckt}
and would break severely the known theoretical constraints on the
odderon amplitude~\cite{Lappi:2016,Berntsen:1991}
\begin{equation}
\left(4-3\mathcal{N}\left(\boldsymbol{r},\,\boldsymbol{b}\right)\right)\mathcal{N}^{3}\left(\boldsymbol{r},\,\boldsymbol{b}\right)-6\left(6-6\mathcal{N}\left(\boldsymbol{r},\,\boldsymbol{b}\right)+\mathcal{N}^{2}\left(\boldsymbol{r},\,\boldsymbol{b}\right)\right)\mathcal{O}^{2}\left(\boldsymbol{r},\,\boldsymbol{b}\right)-3\mathcal{O}^{4}\left(\boldsymbol{r},\,\boldsymbol{b}\right)\ge0.\label{eq:constraint}
\end{equation}
For this reason, we believe that the observed TOTEM-D0 mismatch hardly
can be attributed to the odderon signal. Our findings indicate that
in the framework of chosen parametrizations, the large-$t$ data from
ISR impose more stringent upper limits for odderon amplitude than
the TOTEM-D0 data.  While in all fits we found reasonable values
of $\chi^{2}/{\rm d.o.f.}$, we may observe that the errors of the
parameter $\lambda$ remain elevated, which signals that existing
experimental data constrain poorly the amplitude of odderons. This
is not surprising, given the low statistical significance of the odderons.

\begin{figure}
\begin{tabular}{cc}
\includegraphics[width=8.4cm]{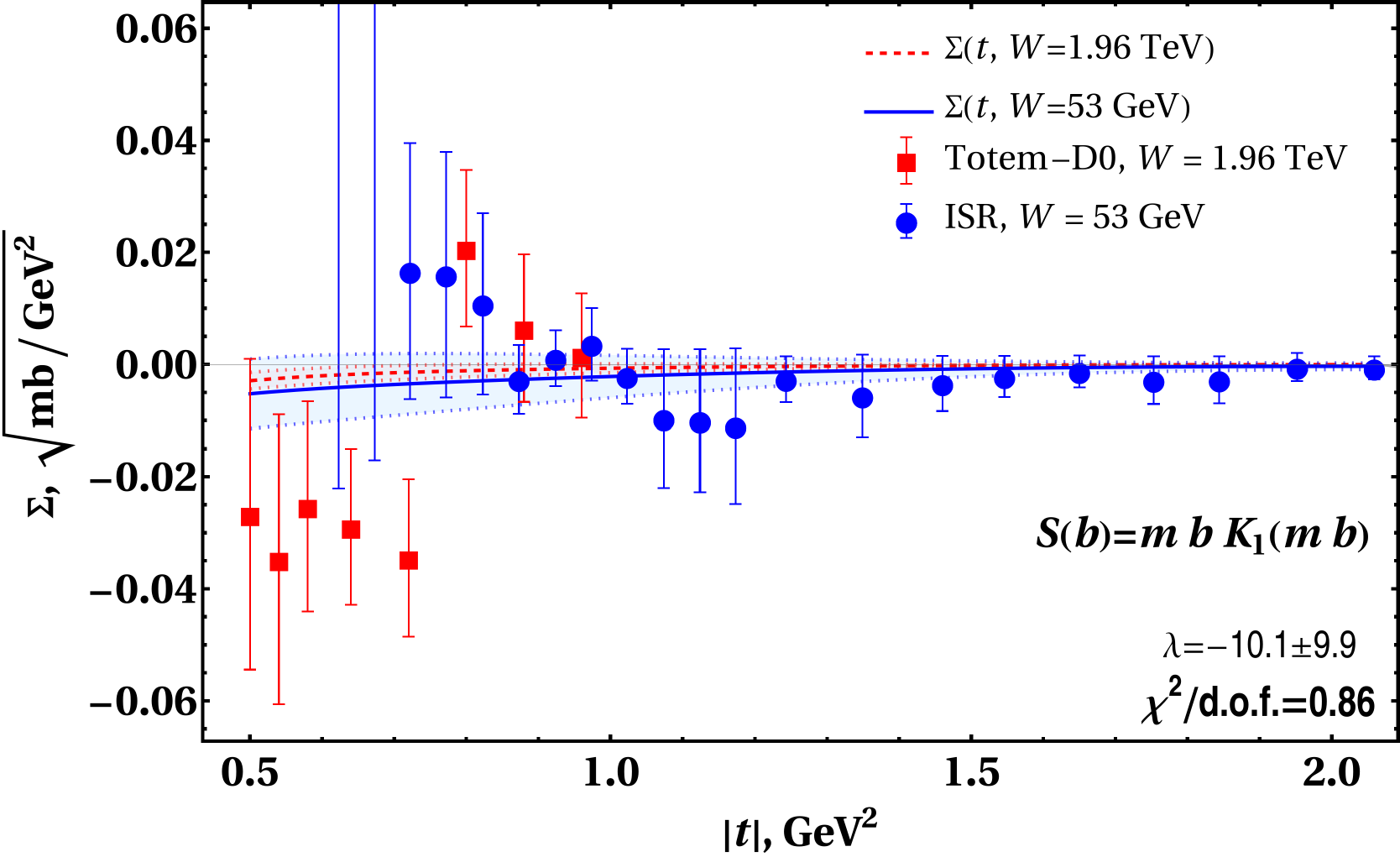} & \includegraphics[width=8.4cm]{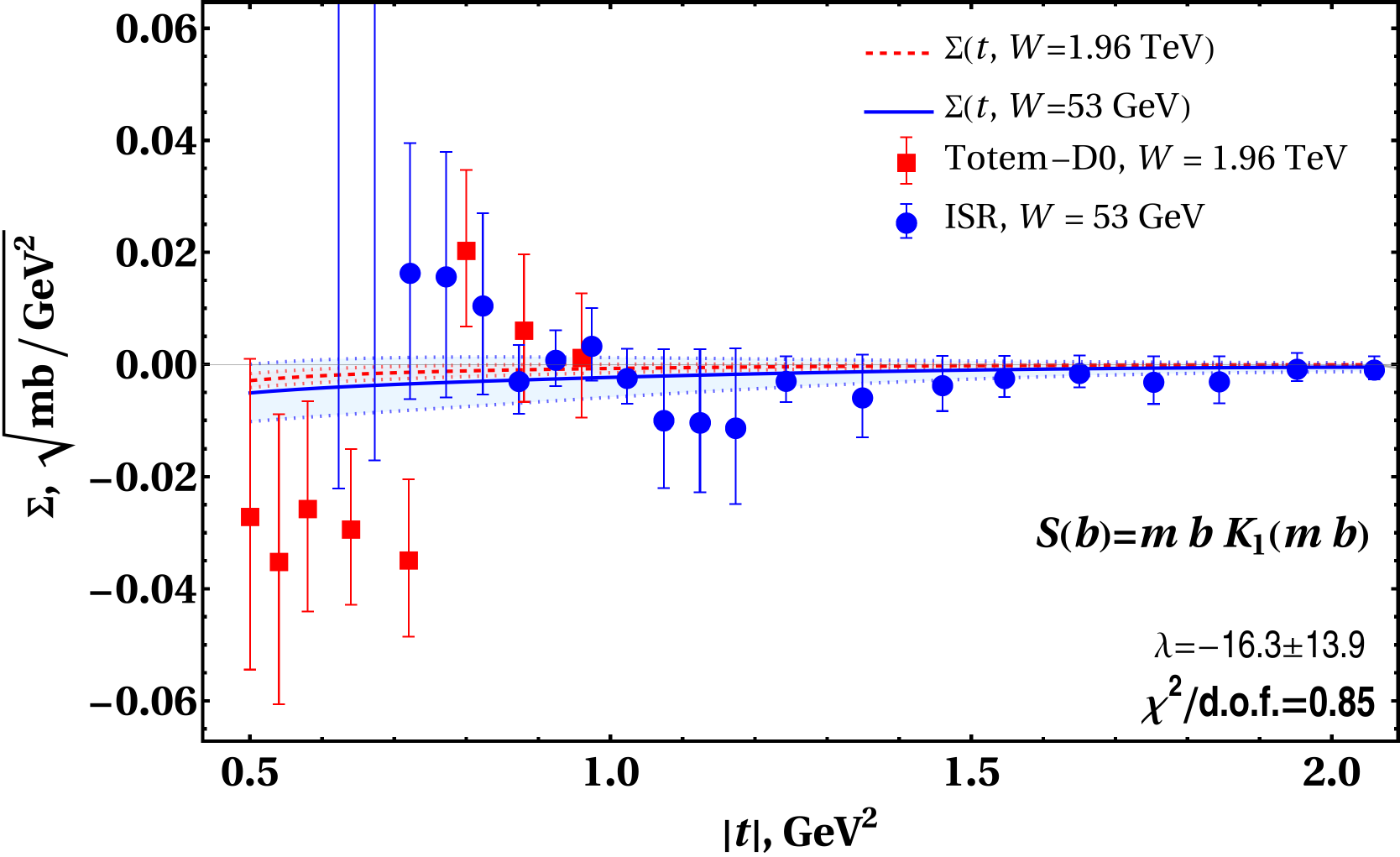}\tabularnewline
\includegraphics[width=8.4cm]{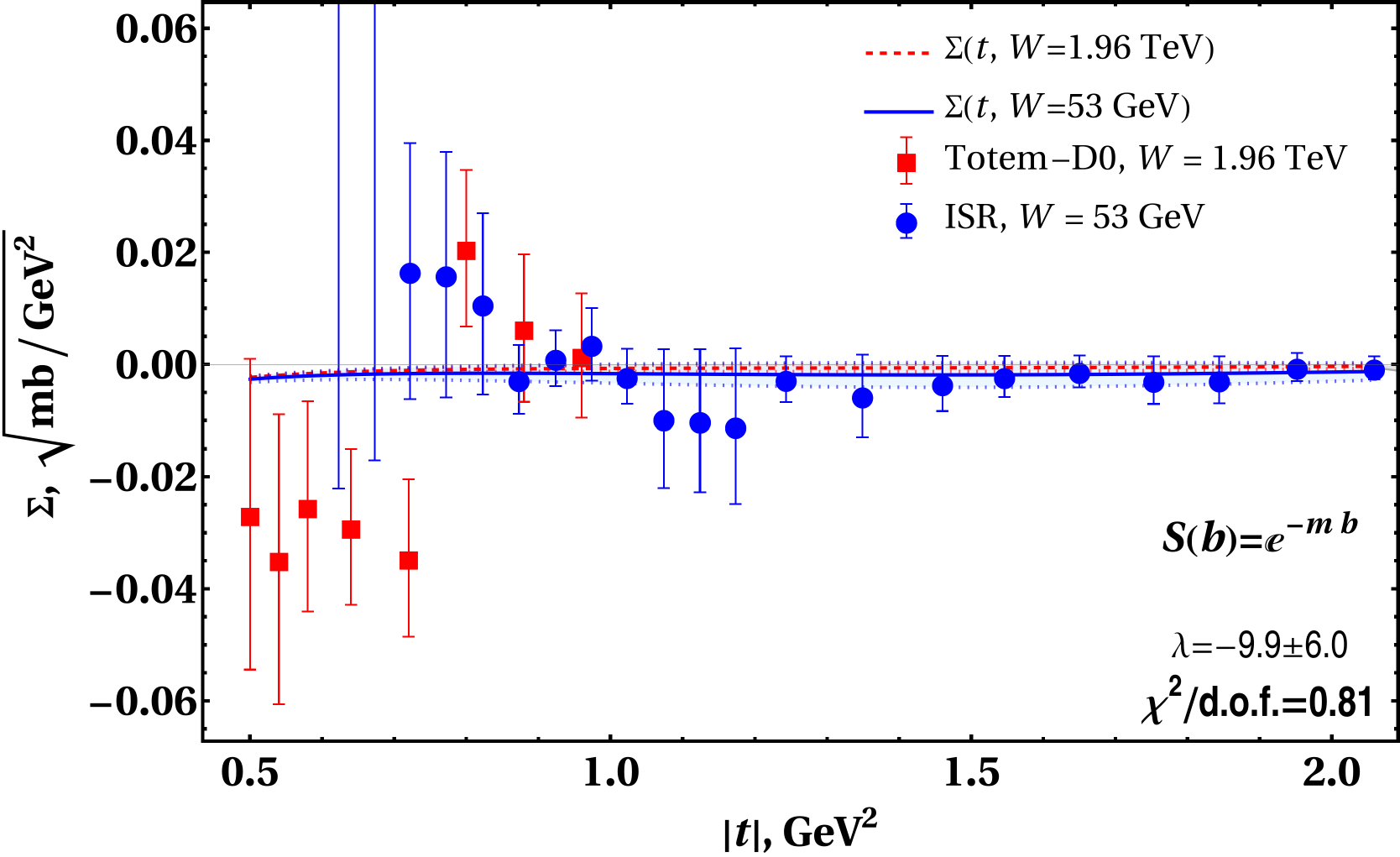} & \includegraphics[width=8.4cm]{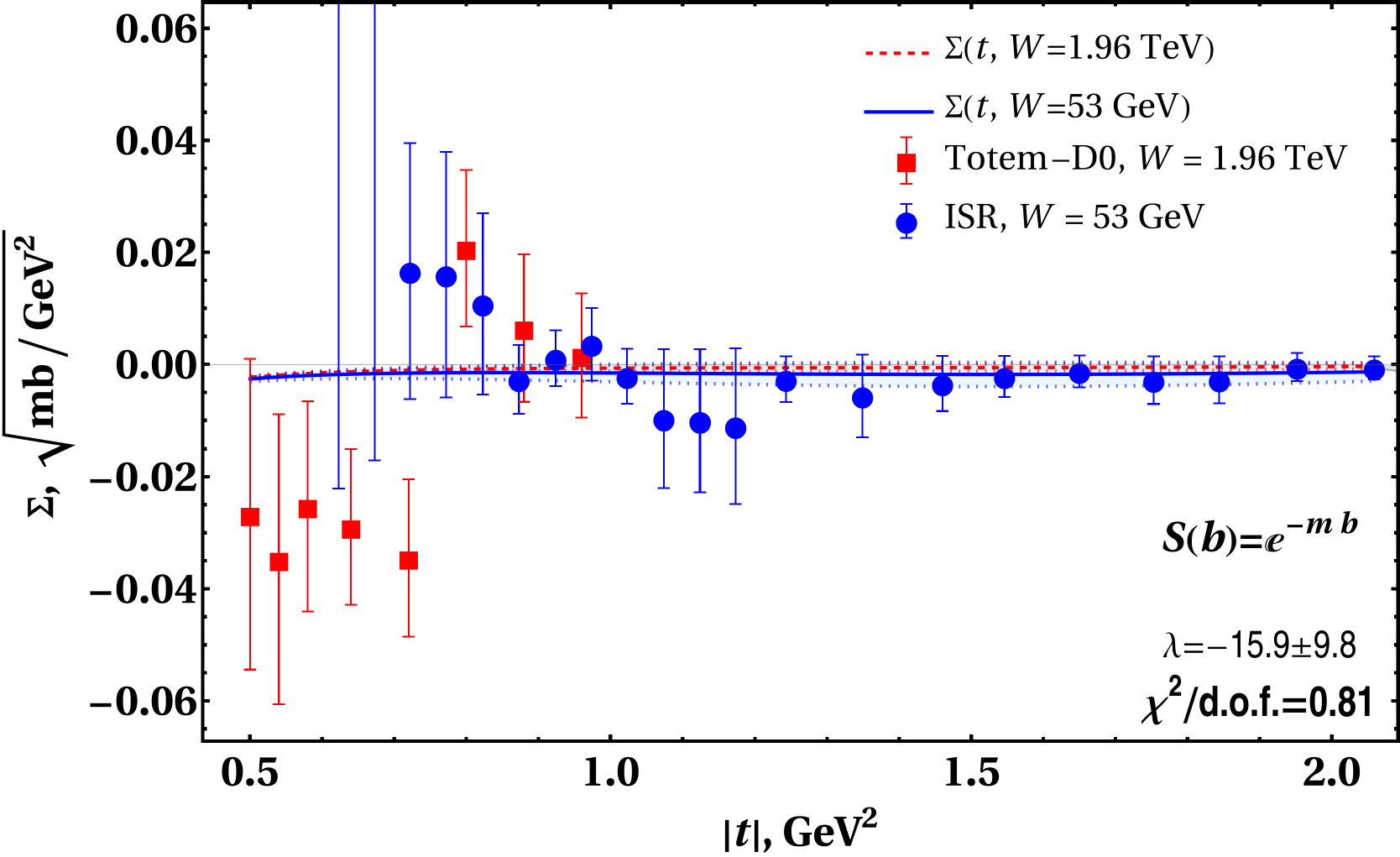}\tabularnewline
\includegraphics[width=8.4cm]{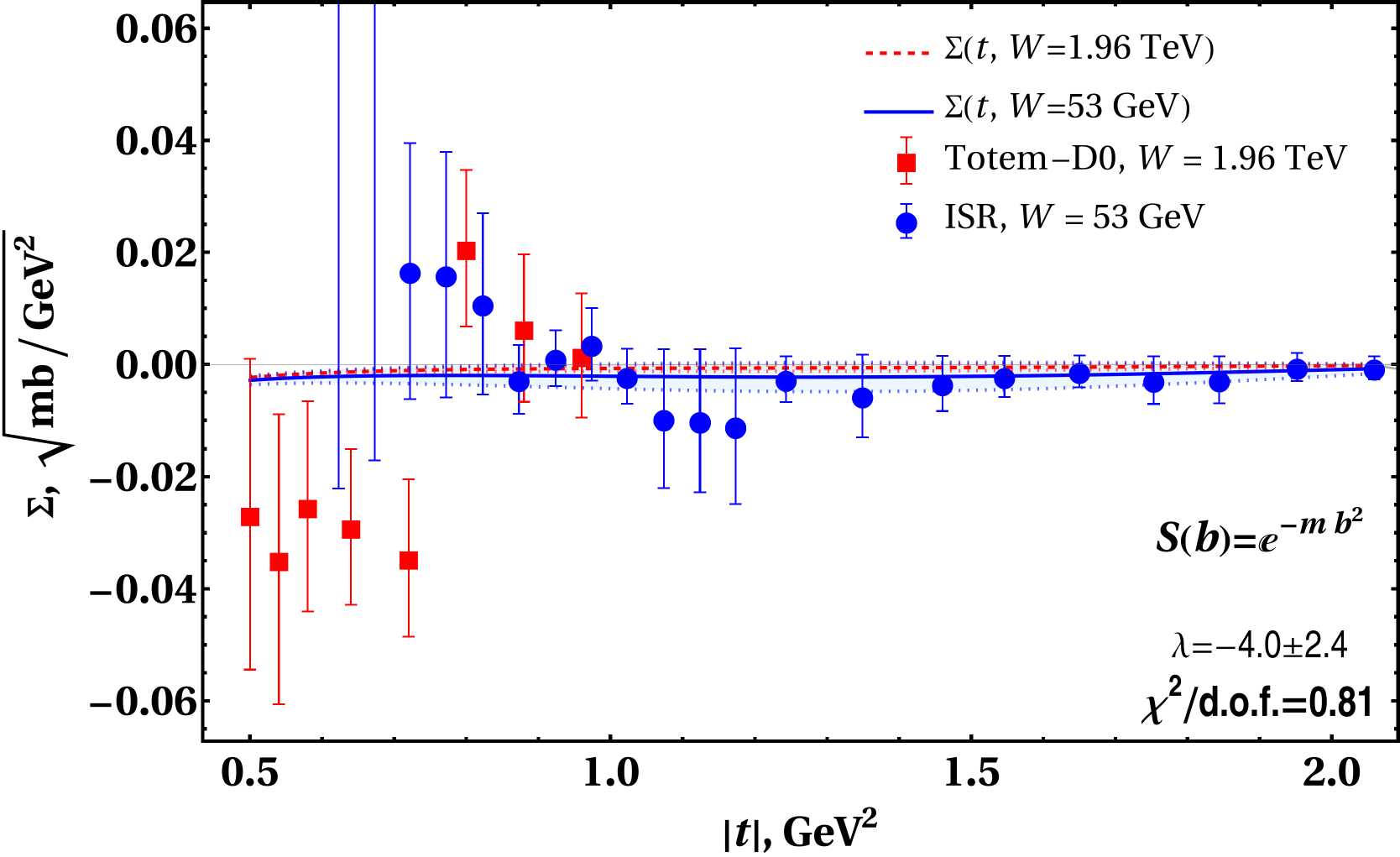} & \includegraphics[width=8.4cm]{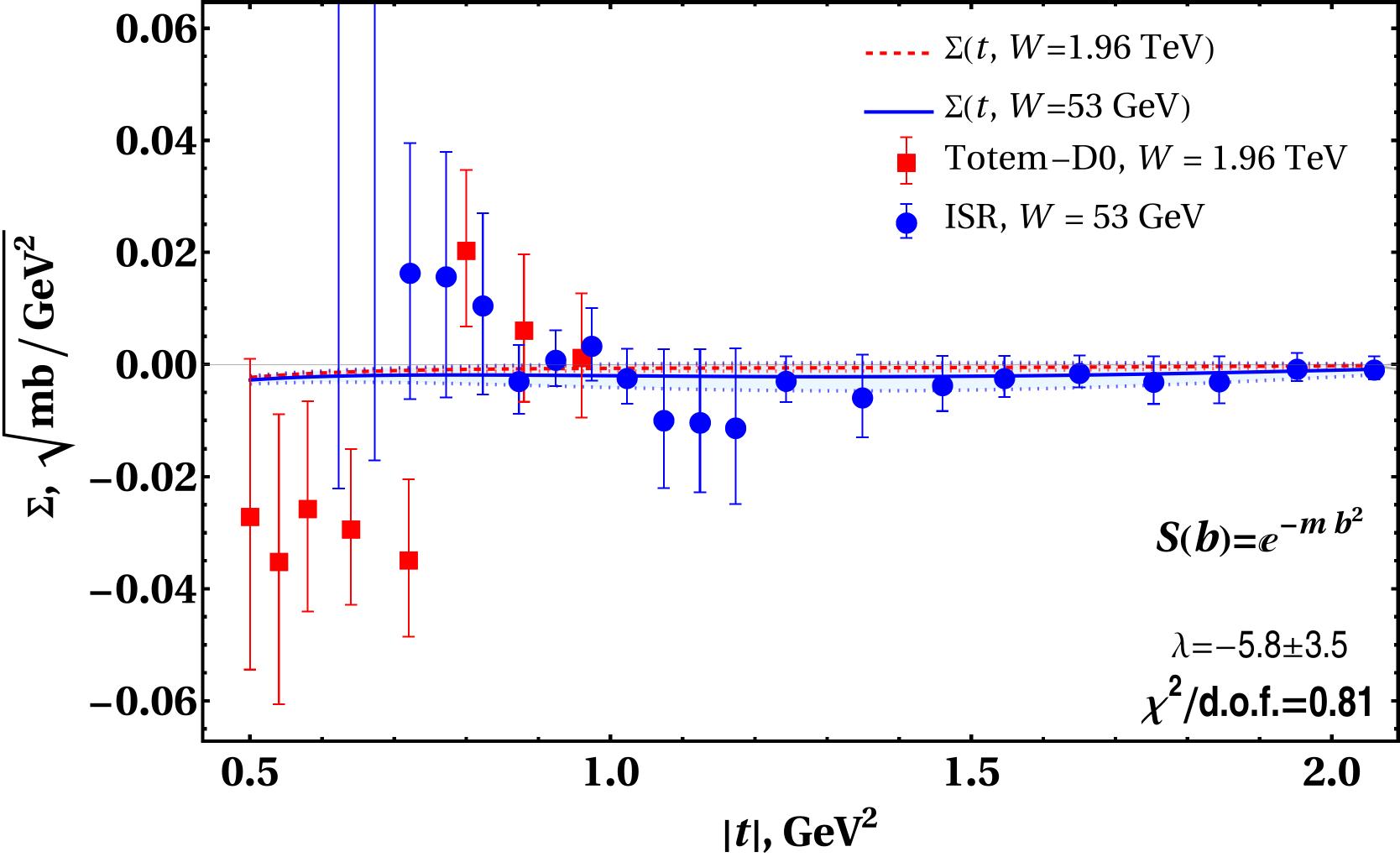}\tabularnewline
\end{tabular}

\caption{(Color online) ~The predictions for the odderon-mediated observable
$\Sigma$ with parametrization~(\ref{OddHatta_b}) and the profiles
$S_{i}(b)$ from the Table~\eqref{tab:profiles} ($m$ is a profile-dependent
constant). The left and right columns correspond to the wave functions~(\ref{eq:dIgauss},\ref{LFWFb}),
respectively. The quoted values of $\chi^{2}/{\rm d.o.f.}$ correspond
to the global fit; the fit of ISR-only data decreases $\chi^{2}/{\rm d.o.f.}$,
however yields nearly the same values of $\lambda$. The points with
error bars (\textquotedblleft experimental data\textquotedblright{}
for the observable $\Sigma$) were derived from ISR~\cite{Breakstone:1985}
and D0-TOTEM experimental data in~\cite{Martynov:2017zjz,D0:2020tig,D0:2012erd,TOTEM:2018psk}.}
\label{fig:odderon_Hatta_separate}
\end{figure}

In the Figure~\eqref{fig:odderon_Hatta_separate} we provided result
of a similar analysis for the parametrization~~(\ref{OddHatta_b}),
assuming that the profile $S(b)$ is given by one of the functions
$S_{1}...S_{3}$ defined in the Table~\eqref{tab:profiles}. For
all possible choices of $S_{i}(b)$, the $t$-dependence of the parameter
$\Sigma$ is very similar to what we obtained earlier using the parametrization~(\ref{eq:oddben}).
As explained earlier, the uncertainty in the choice of the wave function
has only minor effect on the shape of the $t$-dependence and largely
reduces to normalization uncertainty; for this reason the fits with
different wave functions (left and right columns) differ only by the
values of constant $\lambda$ that minimizes $\chi^{2}/{\rm d.o.f.}$

\begin{figure}
\begin{tabular}{cc}
\includegraphics[width=8.4cm]{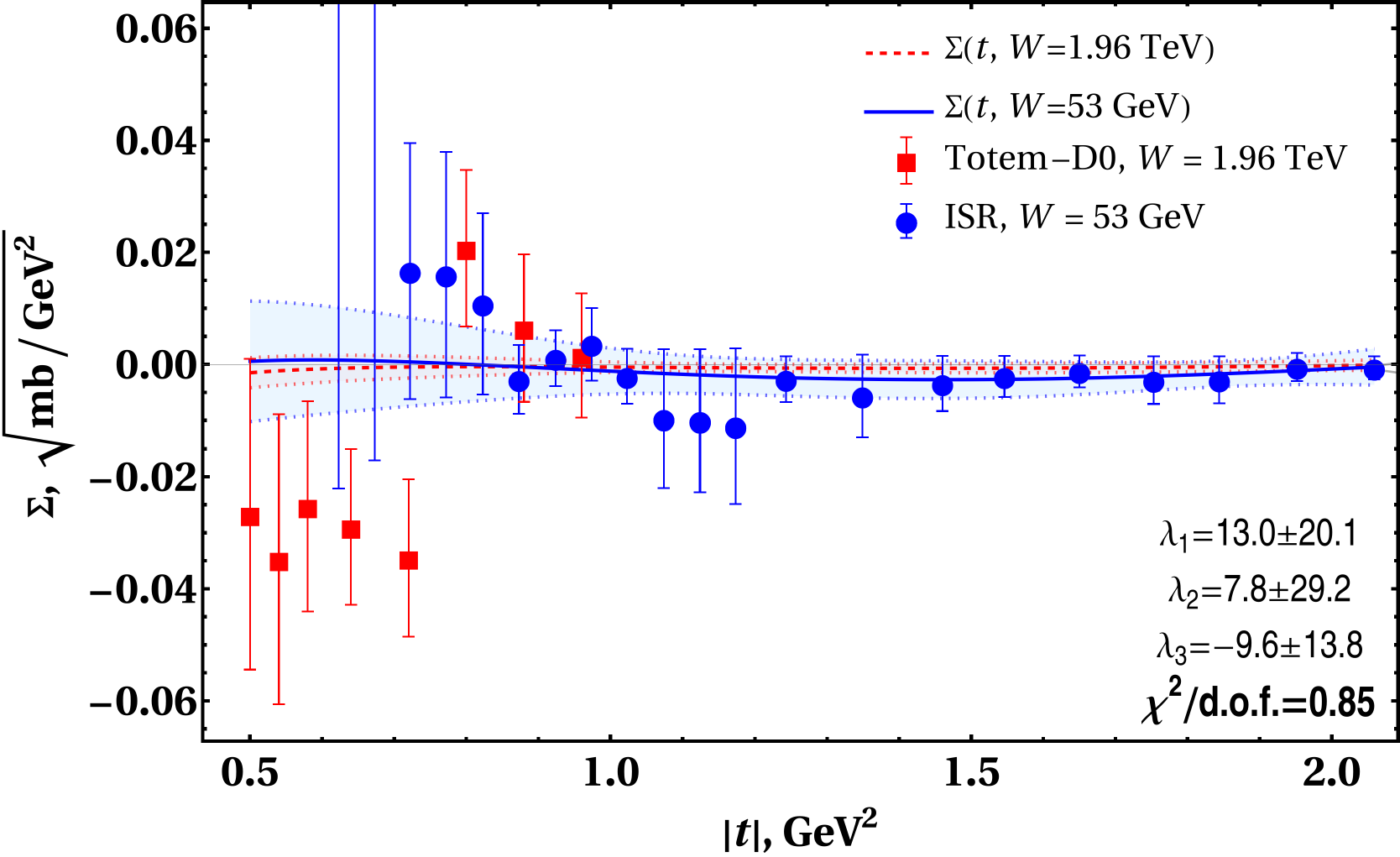} & \includegraphics[width=8.4cm]{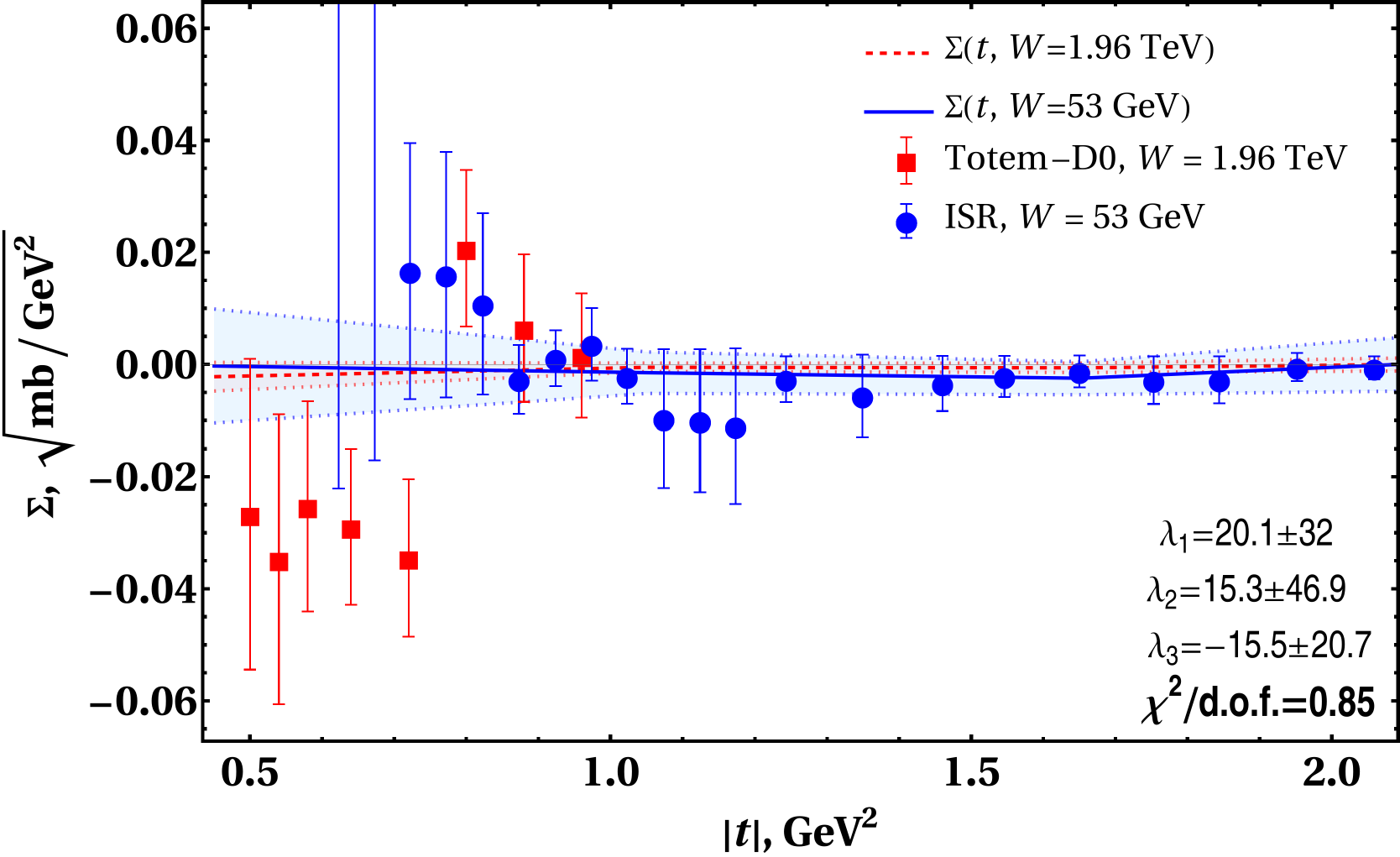}\tabularnewline
\end{tabular}

\caption{The predictions for the odderon-mediated observable $\Sigma$ with
parametrization~(\ref{OddHatta_b}) and profile~(\ref{eq:Linear}).
The left and right columns correspond to the wave functions~(\ref{eq:dIgauss},\ref{LFWFb}),
respectively. The points with error bars (\textquotedblleft experimental\textquotedblright{}
data) for the observable $\Sigma$ were derived from ISR~\cite{Breakstone:1985}
and D0-TOTEM experimental data in~\cite{Martynov:2017zjz,D0:2020tig,D0:2012erd,TOTEM:2018psk}.
The quoted values of $\chi^{2}/{\rm d.o.f.}$ correspond to the global
fit; the fit of ISR-only data decreases $\chi^{2}/{\rm d.o.f.}$ ,
however yields nearly the same values of $\lambda_{n}$.}
\label{fig:odderon_Hatta_all}
\end{figure}

Finally, in the Figure~\ref{fig:odderon_Hatta_all} we present the
results of analysis in which we assumed that the profile $S(b)$ in
the parametrization~~(\ref{OddHatta_b}) may be represented as a
linear superposition of the functions $S_{1}...S_{3}$ defined in
the Table~\eqref{tab:profiles}, namely 
\begin{equation}
S(b)=\sum^{3}_{n=1}\lambda_{n}S_{n}(b),\label{eq:Linear}
\end{equation}
and treated $\lambda_{n}$ as free parameters. This parametrization
is sufficiently flexible to represent any profile that has a maximum
at the center and decreases on the proton's periphery. In view of
the above-mentioned linearity of the BK evolution for odderons in
the limit $\mathcal{O}\ll\mathcal{N}$, each term in the right-hand
side of~(\ref{eq:Linear}) can be evolved independently, thus drastically
simplifying the numerical procedure. Due to similarity of the shapes
of the functions $S_{n}(b)$, this approach suffers from sizable correlation
between different parameters (the corresponding nondiagonal elements
of the correlation matrix are between 0.5 and 1). For this reason,
the fitted curves have enhanced error bands compared to previous fits.
At TOTEM-D0 energy range, the contribution of odderon remain small
due to suppression of its amplitude by BK evolution; for this reason
the inclusion of TOTEM-D0 data increases the $\chi^{2}/{\rm d.o.f.}$,
yet has a very mild influence on the fit parameters.

Although the extracted constraints suggest that the normalization
factor $\lambda$ may be large , we checked that the values of $\lambda\lesssim10$
neither undermine our assumption that $\left|\mathcal{A}_{\mathbb{O}}\right|^{2}$
term in~(\ref{eq:XSecSimplified}) can be neglected, nor leads to
unphysically large cross-section for exclusive $\pi^{0}$ production
at HERA~\cite{H1:2002ckt}.

\section{Conclusions}

\label{sec:Conclusions}In summary, in this manuscript we developed
a formalism which can be used to extract the $C$-odd contributions
to the elastic nucleon-nucleon scattering and constrain phenomenologically
the odderon amplitude $\mathcal{O}(\boldsymbol{r},\,\boldsymbol{b})$
in the Color Glass Condensate framework. In contrast to previous studies,
we constructed the combination of the $pp,\,p\bar{p}$ cross-sections
defined in~(\ref{eq:Razr}) which is directly proportional to the
amplitude mediated by $C$-odd exchanges in the $t$-channel. We also
demonstrated that the smallness of odderon amplitude $(\mathcal{O}\ll\mathcal{N})$
allows to completely decouple the equation for the $C$-even amplitude
$\mathcal{N}$, and thus a global simultaneous fit of the $C$-odd
and $C$-even amplitudes reduces to their sequential extraction from
the experimental data. Since the experimental datasets used to fix
the amplitudes $\mathcal{N}$ and $\mathcal{O}$ have a significantly
different number of events and precision, such sequential procedure
allows to measure more accurately the feeble odderon signal. We demonstrated
that variation of the dipole amplitude within a range allowed by the
experimental data from HERA has only minor influence on the odderon
evolution.

Using popular parametrizations of the wave functions, we found upper
constraints on the magnitude of the odderon amplitude. While our estimates
depend on the choice of the parameter $\rho$ defined in~(\ref{eq:N})
and the wave function of the proton, we found that these nonperturbative
factors affect only the global normalization (at most a factor of
two uncertainty), but almost do not change the \emph{shape} of the
$t$-dependence of the cross-section, which is controlled by the impact
parameter of the odderon. In our phenomenological studies we focused
on the TOTEM-D0 and ISR data, although these data sets are only marginally
consistent with each other in the kinematics $|t|<0.7\,{\rm GeV^{2}}$.
We analyzed two widely used parametrizations of odderons and found
that adding them at face value almost does not change $\chi^{2}/N$
compared to the photon-only contribution. We also tried to vary their
magnitudes (normalizations), treating them as free parameters, however
this only marginally improved description of the experimental data.
In these global fits, the most stringent constraints on odderon amplitude
come from the ISR data at large $|t|>1\,{\rm GeV^{2}}$. While the
obtained values of $\chi^{2}/{\rm d.o.f.}$ are small, due to large
experimental uncertainties, the constraints on the odderon magnitude
are very loose and suggest that the normalization constant $\lambda$
is negative. The inclusion of the TOTEM-D0 data increases the global
$\chi^{2}/{\rm d.o.f.},$ yet has only mild influence on the parameters
because odderon amplitude is suppressed at D0 kinematics by BK evolution.
The attempts to interpret the TOTEM-D0 data as a manifestation of
odderon exchanges would inevitably require a significantly larger
odderon amplitude, violating the known theoretical constraint~(\ref{eq:constraint}).
Furthermore, this would lead to a tension with ISR and HERA data.
Our findings agree with earlier claims in~\cite{Luna:2024cbq} that
TOTEM-D0 signal is too large for odderons and may underestimate certain
systematic errors that affect global normalization of data measured
by different collaborations. Precisely, the authors of~\cite{Luna:2024cbq}
had to introduce the normalization factors $N_{i}$ to reconcile discrepancies
of different experiments and get a consistent global fit.

Through a detailed comparison of multiple saturation scale profiles
and proton wave function parametrizations, we have shown that our
analysis remains robust across a variety of modeling choices, and
thus captures the essential physics of $C$-odd exchange. Unfortunately,
the large errors of ISR and TOTEM-D0 data for $|t|\in\left(0.5,\,0.7\right)\,{\rm GeV^{2}}$,
as well as lack of other experimental evidence in favor of odderons,
does not allow us to make more precise fits~\footnote{While there is plenty of high-precision $pp$ elastic scattering data
from LHC, to the best of our knowledge, at present there is no plans
to study experimentally the $p\bar{p}$ elastic scattering in the
foreseeable future at the same energies. For this reason, all the
attempts to extract the odderon signal from $pp$ scattering at LHC
would require extrapolation to lower energies of older $p\bar{p}$
experiments, and inevitably would suffer from the same problem as
the above-mentioned TOTEM-D0 joint analysis.}. However, the approach developed here lays the groundwork for future
developments and allows to consider on equal footing the odderon contribution
to elastic nucleon-nucleon scattering together with odderon signal
on other channels. Potentially such constraints can be obtained from
spin-sensitive polarized observables~\cite{Hagiwara:2020mqb} and
odderon-mediated quarkonia photoproduction~\cite{Benic:2023ybl}
that can be measured at the future EIC.

\section*{Acknowledgments}

We thank our colleagues at UTFSM university and Eugene Levin for encouraging
discussions. This research was partially supported by ANID CCTVal
CIA250027 (Chile), and ANID grants Fondecyt Regular \textnumero 1251975
and Fondecyt Postdoctoral \textnumero 3230699. I. Z. acknowledges
funding from the Fellowship ANID Beca de Doctorado Nacional No. 21250067
and comprehensive support from the Institute of Physics of PUCV. Y.
G. acknowledges financial support provided by the Universidad de Playa
Ancha de Ciencias de la Educaci\'on through the 2024 Research Seed Grants
Program (Concurso Semilleros de Investigaci\'on 2024), D.E. 0295/2024,
Project Code SEIV 24-24, which significantly contributed to the development
of this research. Powered@NLHPC: This research was partially supported
by the supercomputing infrastructure of the NLHPC (ECM-02).
 \section*{DATA AVAILABILITY}
 The main results of this manuscript are the theoretical expressions
 for amplitudes and are contained within the article. The phenomenological
 estimates for the cross-sections presented in this manuscript are
 also provided as tabulated plaintext data files in Section Supplementary
 Material~\cite{SupplMat} of this manuscript.


\end{document}